# Maturation Trajectories of Cortical Resting-State Networks Depend on the Mediating Frequency Band

S. Khan[1,4,5‡*], J. A. Hashmi[1,4‡], F. Mamashli[1,4], K. Michmizos[1,4], M. G. Kitzbichler[1,4], H. Bharadwaj[1,4], Y. Bekhti[1,4], S. Ganesan[1,4], K. A Garel[1, 4], S. Whitfield-Gabrieli[5], R. L. Gollub[2,4], J. Kong[2,4], L. M. Vaina[4,6], K. D. Rana[6], S. S. Stufflebeam[3,4], M. S. Hämäläinen[3,4], T. Kenet[1,4]

[1]Department of Neurology, MGH, Harvard Medical School, Boston, USA.
[2]Department of Psychiatry MGH, Harvard Medical School, Boston, USA.
[3]Department of Radiology, MGH, Harvard Medical School, Boston, USA.
[4]Athinoula A. Martinos Center for Biomedical Imaging, MGH/HST, Charlestown, USA
[5]McGovern Institute for Brain Research, Massachusetts Institute of Technology, Cambridge, USA
[6]Department of Biomedical Engineering, Boston University, Boston, USA

‡equal contribution

Running Title: Multifaceted Development of Brain Connectivity with Age

***Corresponding author**:
Sheraz Khan, Ph.D.
Athinoula A. Martinos Center for Biomedical Imaging
Massachusetts General Hospital
Harvard Medical School
Massachusetts Institute of Technology
149 13th Street, CNY-2275
Boston, MA-02129, USA
Phone: +1 617-643-5634
Fax:  +1 617-948-5966
E-mail: sheraz@nmr.mgh.harvard.edu

**ABSTRACT**

The functional significance of resting state networks and their abnormal manifestations in psychiatric disorders are firmly established, as is the importance of the cortical rhythms in mediating these networks. Resting state networks are known to undergo substantial reorganization from childhood to adulthood, but whether distinct cortical rhythms, which are generated by separable neural mechanisms and are often manifested abnormally in psychiatric conditions, mediate maturation differentially, remains unknown. Using magnetoencephalography (MEG) to map frequency band specific maturation of resting state networks from age 7 to 29 in 162 participants (31 independent), we found significant changes with age in networks mediated by the beta (13-30Hz) and gamma (31-80Hz) bands. More specifically, gamma band mediated networks followed an expected asymptotic trajectory, but beta band mediated networks followed a linear trajectory. Network integration increased with age in gamma band mediated networks, while local segregation increased with age in beta band mediated networks. Spatially, the hubs that changed in importance with age in the beta band mediated networks had relatively little overlap with those that showed the greatest changes in the gamma band mediated networks. These findings are relevant for our understanding of the neural mechanisms of cortical maturation, in both typical and atypical development.

**1. INTRODUCTION**

Synchronous neuronal activity in the brain gives rise to rhythms, that are known to be functionally significant. These rhythms are commonly divided into five fundamental frequency bands, most commonly classified as delta (1-2 Hz), theta (3-7 Hz), alpha (8-12 Hz), beta (13-30 Hz), and gamma (31-80 Hz) (Buzsáki G 2006). One of the hypothesized roles of these rhythms is in forming neuronal ensembles, or networks, via local and longer-range synchronization, across spatially distributed regions (Fries P 2005; Siegel M et al. 2012; Bastos AM et al. 2015; Fries P 2015). Brain networks that emerge in the absence of any directive task or stimulus, referred to as resting state networks (Raichle ME et al. 2001; Raichle ME 2015), have attracted particular interest due to their consistency across and within individuals. Abnormalities in these networks are also emerging as a hallmark of psychiatric and developmental disorders (Broyd SJ et al. 2009; Toussaint PJ et al. 2014; Kitzbichler MG et al. 2015), further underscoring their functional significance. While resting state networks have been studied extensively using fcMRI (functional connectivity MRI), a technique that relies on the slow hemodynamic signal and thus has a maximal temporal resolution of about 1Hz, studies using high temporal resolution magnetoencephalography (MEG), have confirmed that the five fundamental faster rhythms mediate these networks in non-overlapping patterns (de Pasquale F et al. 2010; Hipp JF et al. 2012)

As part of understanding the function of resting state networks in general, and their role in cognitive development and neurodevelopmental disorders in particular, it is important to map their maturational trajectories, from childhood to adulthood. To date, our knowledge of maturational changes in macro-scale functional networks in the developing brain is largely based on task-free fcMRI studies. Several such studies have shown developmental changes in resting state brain networks, where regions associated with separate networks become connected while closely linked local subnetworks lose some of their connections with maturation (Dosenbach NU et al. 2010; Sato JR et al. 2014; Sato JR et al. 2015). Most of these studies have concluded that network integration, how well different components of the network are connected, increases with maturation, while network segregation, the differentiation of the network into modules, or clusters, decreases with maturation. The spatial distribution of hubs, the most highly connected brain regions, also changes with maturation. Another feature examined in prior studies is the small-world property of brain networks. Small world networks optimize the balance between local and global efficiency. fcMRI studies have not documented a change in the small world property of brain networks with maturation from childhood, around age 7, to adulthood, around age 31 (Fair DA et al. 2009). Network resilience, a measure of the robustness of the network as hubs are removed, which has been used to assess robustness in psychiatric disorders (Lo CY et al. 2015), has been shown to be age dependent in infants (Gao W et al. 2011), but age dependency through maturation has not been studied. It has also been shown that the association between global graph metrics characterizing network properties and the ages of the participants follows an asymptotic growth curve (Dosenbach NU *et al.* 2010).

      While fMRI studies have greatly increased our understanding of the development of resting state networks from childhood to adulthood, the relative temporal coarseness of fcMRI makes it impossible to differentiate maturational trajectories by frequency bands (Hipp JF and M Siegel 2015). Mapping the contributions of distinct frequency bands to maturational trajectories is critical because these rhythms are associated with distinct neurophysiological generators (Uhlhaas PJ et al. 2008; Ronnqvist KC et al. 2013), have been mapped to a multitude of cognitive functions (Harris AZ and JA Gordon 2015), are known to themselves change in power and phase synchrony with maturation (Uhlhaas PJ et al. 2009; Uhlhaas PJ et al. 2010).

      To better understand the contribution of individual rhythms to network maturation, we used MEG, which measures magnetic fields associated with neural currents with millisecond time resolution, and has a spatial resolution on the order of a centimeter (Lin FH, JW Belliveau, et al. 2006). We chose to use graph theory with connectivity measured using envelope correlations (Hipp JF *et al.* 2012) as the core metric, to analyze cortical resting state (relaxed fixation) MEG signals from 131 individuals (64

females), ages 7 to 29, in each of the five fundamental frequency bands. We focused on five well-studied graph theory metrics because the approach is well-suited for studying global network properties also in the functional domain (Bullmore E and O Sporns 2009; Rubinov M and O Sporns 2010; Bullmore E and O Sporns 2012; Misic B et al. 2016; Bassett DS and O Sporns 2017). The results were then validated using similar data from 31 individuals (16 females, ages 21-28) from an independent early adulthood resting state data set (Niso G et al. 2015). The full distribution of participants is shown in Figure S1 in SM. Lastly, to determine the relevance of these graph metrics to the maturation of resting state networks within each frequency band, we used machine learning to quantify the extent to which the MEG derived graph metrics can be used to predict age, similarly to a prior resting state networks study that used fMRI data (Dosenbach NU *et al.* 2010). We then assessed whether the data from the independent dataset fit on the same curves.

## 2. MATERIALS AND METHODS

The analysis stream we followed is illustrated in Figure 1.

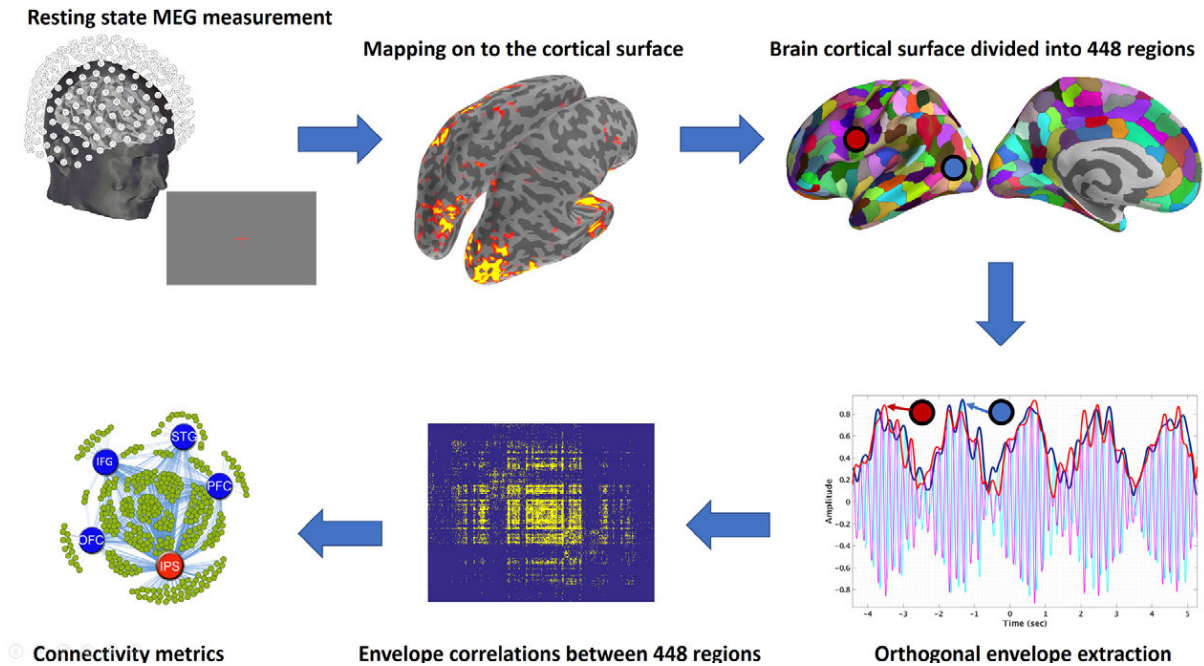

**Figure 1: Schematic illustration of pipeline.** From top left in a clockwise direction: Resting state data are acquired using MEG, and then mapped to the cortical surface. The surface is then divided into regions (parcellated), and envelopes are calculated for each frequency band, in each region. The connectivity between the regions is then computed from the envelopes, and, finally, connectivity metrics are derived.

### 2.1 Experimental Paradigm

The resting state paradigm consisted of a red fixation cross at the center of the screen, presented for 5 minutes continuously, while participants were seated and instructed to fixate on the cross. The fixation stimulus was generated and presented using the psychophysics toolbox (Brainard, 1997; Pelli, 1997), and projected through an opening in the wall onto a back-projection screen placed 100 cm in front of the participant, inside a magnetically shielded room.

### 2.2 Participants

#### *2.2.1 Massachusetts General Hospital (MGH) Based Participants*

Our primary data were collected from 145 healthy subjects, ages 7-29, at MGH. Due to excessive motion, data from 14 subjects were discarded. Because datasets from different MGH based studies were combined here, no uniform behavioral measures were available across all participants. IQ measured with the Kaufman Brief Intelligence Test – II (Kaufman and Kaufman, 2004) was available for 68 of the participants. Within this subgroup, no significant change in IQ with age was observed (Figure S2), as expected, given that IQ is normalized by age. All the studies that were pooled for this analysis screened for typical development and health, but the approach varied. The full age and gender distribution of the participants is shown in Figure S1-A.

### *2.2.2 OMEGA Project Participants (McGill University)*

To test our results on an independent dataset, resting-state MEG scans from 31 additional young adult participants (ages 21-28) were obtained from the OMEGA project (Niso G *et al.* 2015), and chosen by order with gender matching to the MGH cohort in that same age range, subject to age restrictions. Note that the OMEGA project spans ages 21-75. While we would have liked to test our results on data from younger subjects, no pediatric MEG resting state data are currently openly available, so this was not possible. The age and gender distributions of the participants are shown in Figure S1-B.

## 2.3 MRI/MEG Data Acquisition

### *2.3.1 MRI Data Acquisition and Processing*

T1-weighted, high resolution MPRAGE (Magnetization Prepared Rapid Gradient Echo) structural images were acquired on either a 1.5T or a 3.0-T Siemens Trio whole-body MRI (magnetic resonance) scanner (Siemens Medical Systems) using either 12 channels or 32 channel head coil at MGH and MIT. The structural data was preprocessed using FreeSurfer (Dale AM et al. 1999; Fischl B et al. 1999). After correcting for topological defects, cortical surfaces were triangulated with dense meshes with ~130,000 vertices in each hemisphere. To expose the sulci in the visualization of cortical data, we used the inflated surfaces computed by FreeSurfer. For the 31 OMEGA dataset participants, templates were constructed from age and gender matched subjects from our data. The same processing steps are followed on this template MRIs.

### 2.3.2 MEG Data Acquisition and Processing

MEG data were acquired inside a magnetically shielded room (Khan S and D Cohen 2013) using a whole-head Elekta Neuromag VectorView system composed of 306 sensors arranged in 102 triplets of two orthogonal planar gradiometers and one magnetometer. The signals were filtered between 0.1 Hz and 200 Hz and sampled at 600 Hz. To allow co-registration of the MEG and MRI data, the locations of three fiduciary points (nasion and auricular points) that define a head-based coordinate system, a set of points from the head surface, and the locations of the four HPI coils were determined using a Fastrak digitizer (Polhemus Inc., Colchester, VT) integrated with the VectorView system. ECG as well as Horizontal (HEOG) and Vertical electro-oculogram (VEOG) signals were recorded. The position and orientation of the head with respect to the MEG sensor array were recorded continuously throughout the session with the help of four head position indicator (HPI) coils (Cheour M et al. 2004). At this stage, we monitored the continuous head position, blinks, and eye movements in real time, and the session was restarted if excessive noise due to the subject's eyes or head movement is recorded. In particular, the subjects and head coils position were carefully visually monitored continusouly, and the session was also restarted if any slouching in the seated position was observed. Pillows, cushions, and blankets were fitted to each individual to address slouching, and readjusted as needed if any slouching was observed. In addition to the human resting state data, 5 min of data from the room void of a subject were recorded before each session for noise estimation purposes.

The acquisition parameters are published elsewhere for the 31-subjects from the OMEGA dataset (Niso G *et al.* 2015). The OMEGA CTF files were converted into a .fif file using mne_ctf2fif function from MNE (Gramfort A et al. 2014). After this conversion, the subsequent processing was the same as for the Vector View data.

**2.4 Noise suppression and motion correction**

The data were spatially filtered using the signal space separation (SSS) method (Taulu S et al. 2004; Taulu S and J Simola 2006) with Elekta Neuromag Maxfilter software to suppress noise generated by sources outside the brain. Since shielded room at MGH is three layers and we have exclusion criteria for subject having dental artifact, only SSS is applied and it temporal extension tSSS was not used. This procedure also corrects for head motion using the continuous head position data described in the previous section.

Since SSS is only available for Electa MEG systems, it was not applied for OMEGA subjects, where data were collected with a CTF MEG system. The heartbeats were identified using in-house MATLAB code modified from QRS detector in BioSig (Vidaurre C et al. 2011). Subsequently, a signal-space projection (SSP) operator was created separately for magnetometers and gradiometers using the Singular Value Decomposition (SVD) of the concatenated data segments containing the QRS complexes as well as separately identified eye blinks (Nolte G and MS Hämäläinen 2001). Data were also low-pass filtered at 144 Hz to eliminate the HPI coil excitation signals.

**2.5 Cortical Space Analysis**

*2.5.1 Mapping MEG Data onto Cortical Space*

The dense triangulation of the folded cortical surface provided by FreeSurfer was decimated to a grid of 10,242 dipoles per hemisphere, corresponding to a spacing of approximately 3 mm between adjacent source locations. To compute the forward solution, a boundary-element model with a single compartment bounded by the inner surface of the skull was assumed (Hämäläinen MS and J Sarvas 1989). The watershed algorithm in FreeSurfer was used to generate the inner skull surface triangulations from the MRI scans of each participant. The current distribution was estimated using the regularized minimum-norm estimate (MNE) by fixing the source orientation to be perpendicular to the cortex. The regularized (regularization = 0.1) noise covariance matrix that was used to calculate the inverse operator was estimated from data acquired in the absence of a subject before each session.

This approach has been validated using intracranial measurements (Dale AM et al. 2000). To reduce the bias of the MNEs toward superficial currents, we incorporated depth weighting by adjusting the source covariance matrix, which has been shown to increase spatial specificity (Lin FH, T Witzel, et al. 2006). All forward and inverse calculations were done using MNE-C (Gramfort A *et al.* 2014).

### 2.5.2 Correlation between age, and the norms of the columns of the gain matrix

We also examined the correlation between age, and the norms of the columns of the gain matrix, separately for magnetometers and gradiometers. The magnetometers showed no significant correlation with age. The gradiometers showed a small correlation in the frontal pole, in a region that was not identified as a significant hub in our results (see Figure S3). This is most likely due to the different in head size between the youngest participants and the oldest ones. To minimize the effect of these differences, we use both magnetometer and gradiometer data in our source estimation procedure.

### *2.5.3 Cortical Parcellation (Labels)*

FreeSurfer was used to automatically divide the cortex into 72 regions (Fischl B et al. 2004). After discarding "medial wall" and "corpus callosum", these regions were further divided in to a total of *N*=448 cortical "labels (Figure S4)", so that each label covers a similar area again using FreeSurfer. We employed this high-resolution parcellation scheme because cortical surface is very convoluted and averaging across a large label, which crosses multiple sulci and gyri, can result in signal cancellation across the parcel. Lastly, a high-resolution parcellation also reduces the dependence of the results on the specific selection of the parcels.

We also checked that field spread (spatial signal leakage) between labels is not impacted by age by examining the correlation between age and the mean cross-talk across labels (Hauk O et al. 2011). The results showed no correlation (Figure S5).

### *2.5.4 Averaging the time series across a label*

Owing to the ambiguity of individual vertex (dipole) orientations, these time series were not averaged directly but first aligned with the dominant component of the multivariate set of time series before calculating the label mean. In order to align sign of the time series across vertices, we used SVD of the data $X^T = U\Sigma W^T$. The sign of the dot product between the first left singular vector $U$ and all other time-series in a label was computed. If this sign was negative, we inverted the time-series before averaging.

To verify that the label time series are meaningful, we computed the Power Spectral Density (PSD, see Spectral Density, below) for occipital, frontal, parietal, and temporal cortical regions within each age group, see Figure S6.

## 2.6 Time Series Analysis

### *2.6.1 Filtering and Hilbert transform*

The time series were band-pass filtered and down sampled for faster processing, while making sure that the sampling frequency was maintained at $f_s > 3f_{hi}$ (obeying the Nyquist theorem and avoiding aliasing artifacts). The chosen frequency bands were delta (1-4 Hz), theta (4-8 Hz), alpha (8-12 Hz), beta (13-30 Hz), and gamma (31-80 Hz). The line frequency at 60 Hz was removed with a notch filter of bandwidth 1 Hz. Hilbert transform was then performed on this band pass data.

### 2.6.2 Hilbert Transform

For each individual frequency band the analytic signal $\hat{X}(t)$ was calculated by combining the original time series with its Hilbert transform into a complex time series:

$$\hat{X}(t) = x(t) + j\,\mathcal{H}[x(t)}] --- (1)$$

The resulting time series $\hat{X}(t)$ can be seen as a rotating vector in the complex plane whose length corresponds to the envelope of the original time series $x(t)$ and whose phase grows according to the dominant frequency. Figure 1, step 4 shows an example of a modulated envelope on the top of the band pass data (carrier). An example of envelope PSD for the gamma frequency band is shown in Figure S7.

At this stage further artifact cleaning was performed as follows: signal spikes where the amplitude is higher than 5σ over the course were identified and dropped over a width of 5 periods.

To remove the effect of microsaccades, HEOG and VEOG channels were filtered at a pass-band of 31–80 Hz. The envelope was then calculated for the filtered signals and averaged to get REOG. Peaks exceeding three standard deviations above the mean calculated over the whole-time course, were identified and the corresponding periods were discarded from subsequent analysis.

Head movement recordings from the HPI coils were used to drop these one second blocks where the average head movement exceeded 1.7 mm/s (empirical threshold). The amount of data lost through cleaning was well below 10% and did not differed significantly with age.

### 2.6.3 Orthogonal correlations

We used a method based upon envelope correlations to reliably estimate synchronicity between different cortical labels (Colclough G et al. 2016). In contrast to phase-based connectivity metrics envelope correlations measure how the amplitude of an envelope within a frequency band is synchronously modulated over time across distinct cortical regions, as illustrated in the fourth panel of Figure 1. Previous studies (humans and primates) have demonstrated the validity and functional significance of these synchronous envelope amplitude modulations (Brookes MJ et al. 2011; Vidal JR et al. 2012; Wang L et al. 2012; Brookes MJ et al. 2016; Colclough G *et al.* 2016) for both oscillatory and broadband signals.

To address the field-spread problem associated with MEG data (Sekihara K et al. 2011), we used the previously described orthogonal (Hipp JF *et al.* 2012) variation of envelope correlation metric. This method requires any two putatively dependent signals to have non-zero lag and is thus insensitive to the zero-lag correlations stemming from field-spread.

Mathematically, the connectivity between two complex signals $\hat{X}$ and $\hat{Y}$ is calculated by "orthogonalizing" one signal with respect to the other $\hat{Y}(t,f) \rightarrow \hat{Y}_{\perp X}(t,f)$, and subsequently taking the Pearson correlation between their envelopes. This is done in both directions and the two results are averaged to give the final connectivity measure $C_\perp(\hat{X}, \hat{Y}; t, f)$.

$$\hat{Y}_{\perp X}(t,f) = \Im\left(\hat{Y}(t,f) \frac{\hat{X}^\dagger(t,f)}{|\hat{X}(t,f)|}\right) \hat{e}_{\perp X}(t,f) \quad ---(2)$$

$$C_\perp(\hat{X}, \hat{Y}; t, f) = \frac{Corr(|\hat{X}|, |\hat{Y}_{\perp X}|) + Corr(|\hat{Y}|, |\hat{X}_{\perp Y}|)}{2} \quad --(3)$$

Due to the slow time course of these envelopes and to ensure enough independent samples are available in the correlation window (Hipp JF *et al.* 2012), we calculated the orthogonal connectivity using an overlapping sliding window of 30 seconds with a stride of $1/8$ of the window size. Figure S7 demonstrate this method on gamma envelopes.

Lastly, note that this method does not address the problem of spurious correlations appearing due to source spread (Sekihara K *et al.* 2011), which is difficult to address. Therefore, spurious correlations coupled with the use of high resolution cortical parcellation can skew the topological metrics used in this study. Please see (Palva JM et al. 2017) for a comprehensive and detailed discussion of these confounds.

**2.7 The Connectivity and Adjacency Matrices**

As a starting point for calculating the graph theoretic metrics we used the connectivity matrix, which contained the orthogonal correlations between all $N \times N$ node pairs and at each time window. A separate matrix was computed for each frequency band. The final result of the processing pipeline is a connectivity array of dimension $N \times N \times N_{\text{Time}} \times N_{\text{Bands}}$ for each subject. In order to increase signal to noise, we collapsed the connectivity array along the temporal dimension by taking the median of each pairwise orthogonal correlation across time windows.

Thresholding and binarizing the connectivity matrix results in the adjacency matrix $\mathbb{A}$.

We used a threshold proportional scheme to retain a given proportion of the strongest connectivity matrix entries in $\mathbb{A}$. Specifically, adjacency matrix $\mathbb{A}$, were constructed using a fixed cost threshold, ensuring that the density or number of connections of the network is equated across all individuals and age groups. Cost is a measure of the percentage of connections for each label in relation to all connections of the network. Since the total number of connections is the same for all participants, and is determined by the number of nodes being considered, the use of a fixed cost, i.e. fixed percentage threshold, allows for exactly equal numbers of connections across participants. This is important to ensure graph metrics can be compared across all individuals and age groups. As there is no rationale for using a cost threshold, therefore we compared graph network properties for a wide range of cost, we used thresholding range from 5% to 30% at increments of 5%. For the graph metrics to be reliable, it should be consistent over range of thresholds. Please see Table S1 for example of consistency of our results. We also checked that important hubs for all frequency bands are in line with previous published studies (Brookes MJ *et al.* 2011; Hipp JF *et al.* 2012; Colclough G *et al.* 2016) as shown in Figure S8.

The adjacency matrix $\mathbb{A}$ defines a graph $\mathcal{G}$ in the form of pairs of nodes that are connected by an edge. Thus, $\mathbb{A}$ is defined such that its binary elements $\mathbb{A}_{ij}$ are either 1 or 0 depending whether the edge $e_{ij}$ between nodes $v_i$ and $v_j$ exists or not:

$$\mathbb{A}_{ij} = \begin{cases} 1 & \text{if } \exists\ e_{ij} \\ 0 & \text{if } \nexists\ e_{ij} \end{cases}$$

In addition, we also computed a weighted adjacency matrix which preserves the correlation values above the same thresholds. A more comprehensive description can be found elsewhere (Watts DJ and SH Strogatz 1998). For Figures 3B,D and 4B,D, the original adjacency matrices were averaged within age-groups and then thresholded for visualization purposes.

### *2.7.1 Path length*

The average shortest path length between all pairs of nodes was calculated as:

$$L = \frac{1}{n(n-1)} \sum_{i \neq j; v_i, v_j \in \mathcal{G}} d_{ij} \quad --- (4)$$

where the topological distance $d_{ij}$ between nodes $v_i$ and $v_j$ is defined as the minimum number of edges one has to traverse in order to get from one node to the other

$$d_{ij} = \min\{n | \mathbb{A}^n[i,j] \neq 0\}$$

where $\mathbb{A}^n$ denotes the $n$th power of the adjacency matrix $\mathbb{A}$ and $i$ and $j$ are row and column indices of the resulting matrix.

### *2.7.2 Degree*

The degree of a node $v_i$ in a Graph $\mathcal{G}$ is defined as

$$D_i = \sum_{j=1, j \neq i}^{n} e_{ij} \quad --- (5)$$

where $e_{ij}$ is the $i$ th row and $j$ th column edge of adjacency matrix $\mathbb{A}$.

The degree maps, averaged across all participants in the adult group (ages 22-29), for all bands, are shown in Figure S8.

We also computed weighted degree (unthresholded mean connectivity with respect to age, also known as mean node strength) for alpha and beta (Figure S9), and, as expected, the results are in line with prior findings showing increase in overall connectivity with age (Schäfer CB et al. 2014).

### *2.7.3 Clustering coefficient*

The local clustering coefficient in the neighborhood of vertex $v_i$ is defined as the ratio of actual and maximally possible edges in the graph $\mathcal{G}_i$, which is equivalent to the graph density of $\mathcal{G}_i$:

$$C_i = \frac{2|\{e_{jk}\}|}{k_i(k_i - 1)} : v_j, v_k \in \mathcal{G}_i \quad --- (6)$$

### *2.7.4 Global and local efficiencies*

Global efficiency measures the efficiency of information transfer through the entire network, and is assessed by mean path length. While the concept of path length is intuitive in anatomical networks, it is also relevant for functional networks, since a particular functional connection may travel different anatomical paths, and while the correspondence between the two is generally high, it is not necessarily identical (Bullmore E and O Sporns 2009; Misic B *et al.* 2016; Bassett DS and O Sporns 2017). Local efficiency is related to the clustering of a network, i.e. the extent to which nearest neighbors are interconnected. Thus, it assesses the efficiency of connectivity over adjacent brain regions.

The average global efficiency of information transfer in graph $\mathcal{G}$ having $n$ nodes can be calculated from the inverse of the edge distances $d_{i,j}$

$$E_{\text{glob}} = E(\mathcal{G}) = \frac{1}{n(n-1)} \sum_{i \neq j; v_i, v_j \in \mathcal{G}} \frac{1}{d_{ij}} \ ---(7)$$

The quantity above is a measure of the global efficiency of information transfer for the whole graph $\mathcal{G}$. There is also a local efficiency for each vertex $v_i$ measuring how efficiently its neighbours can communicate when vertex $v_i$ is removed. If the subgraph of all neighbors of $v_i$ is denoted by $\mathcal{G}_i$, then its local efficiency $E(\mathcal{G}_i)$ is approximately equivalent to the clustering coefficient $C_i$ (Achard S and E Bullmore 2007).

$$E_{\text{loc}} = \frac{1}{n} \sum_{v_i \in \mathcal{G}} E(\mathcal{G}_i) \ ---(8)$$

We also computed weighted analogues of local and global efficiencies which used $C_\perp$-weighted edge distances $d_{i,j}$. This weighted analogue shows the same trend as a function of age as the unweighted one, please see Figure 2 and Figure S10.

### *2.7.5 Small World*

Small world property is a measure of optimization of the balance between short and long-range connections (Bassett DS and E Bullmore 2006). When graph $\mathcal{G}$ that provides optimal balance between local and global information transfer, it is called a small-world graph. The small-worldness of a network is often characterized by two key metrics: the clustering coefficient $C$ and the characteristic path length $L$. To evaluate the small-world topology of brain networks, these topological parameters must be benchmarked against corresponding mean values of a null random graph. A network $\mathcal{G}$ is is a small-world network if it has a similar path length but greater clustering of nodes than an equivalent Erdös-Rényi (E–R) random graph $\mathcal{G}_{rand}$:

$$SW = \frac{C/C_{rand}}{L/L_{rand}} \geq 1 \quad ---(9)$$

where $C_{rand}$ and $L_{rand}$ are the mean clustering coefficient and the characteristic path length of the equivalent $\mathcal{G}_{rand}$ graph.

### *2.7.6 Betweenness Centrality*

Betweenness centrality pertains to individual nodes in the network, rather than the network as a whole, and assesses how many of the shortest paths between all other node pairs in the network pass through that node. Nodes with high betweenness centrality (hubs) are therefore more important for overall network efficiency.

The betweenness centrality of node *i* is defined as:

$$b_i = \sum_{m \neq i \neq n \in G} \frac{\sigma_{mn}(i)}{\sigma_{mn}} \quad ---(10)$$

where $\sigma_{mn}$ is the total number of shortest paths (paths with the shortest path length) from node *m* to node *n*, and $\sigma_{mn}(i)$ is the number of shortest paths from node *m* to node *n* that pass through node *i*. Betweenness centrality of a node thus reflects the control and influence of that node on other nodes. Nodes with high betweenness centrality have a high impact on information transferal and collaboration between disparate sub-networks.

*2.7.7 Resilience*

Resilience measures the robustness of the network if the most heavily connected nodes (hubs) are removed. This measure is inversely related to small world property (Peng G-S et al. 2016). We chose this measure because it has been studied, mostly using fMRI, in the context of psychiatric disorders, where multiple hubs might be functioning abnormally (Achard S et al. 2006; Lo CY *et al.* 2015). It has also been shown that greater resilience in a functionally derived task-driven network is associated with greater inhibitory control cognitively (Spielberg JM et al. 2015), a function that is often impaired in neurodevelopmental and psychiatric disorders. Importantly, the measure incorporates network topology in conjunction with the spatial distribution of hubs, because it takes the degree, i.e. the number of connections, of individual nodes into account.

Resilience quantifies the Graph $\mathcal{G}$'s robustness to targeted or random attacks. Targeted attacks remove nodes in the descending order of degree. At each attack, global efficiency is computed. Robustness is defined as the ratio of the original efficiency with efficiency calculated after attack. This process is repeated until all nodes are removed. Graph where the connectivity probability follows a power law distribution (i.e. scale free networks) are very robust in the face of random failures. This property of networks mathematically described by a power law function $p(k) \propto k^{-y}$ where $p(k)$ is the probability of a node having k links is and y is the exponent. When plotted in a log-log plot, this relationship follows a straight line with slope *-y*, i.e. resilience is the slope of the degree distribution.

**2.8. Spectral Analysis**

The Power Spectral Density (PSD) (0.1 to 80Hz, logarithmically distributed) in Figure S6 was computed on single trial data from each label using the multitaper method (MTM) based on discrete prolate spheroidal sequences (Slepian sequences) taper(Thomson DJ 1982) with 1.5 Hz smoothing as implemented in MNE-Python. PSD (0.033 to 1Hz, logarithmically distributed) in Figure S7 of gamma envelope was also computed using multitaper method with spectral smoothing of 0.02 Hz.

**2.9. Correlation**

To evaluate the relationship between a network quantity and age, we used Spearman correlation (degree of freedom = 129). The p-values were computed after correcting for multiple comparisons across the correction space of frequency bands, thresholds, and graph metrics by controlling for family-wise error rate using maximum statistics through permutation testing (Groppe DM et al. 2011).

The corrected p-values are shown in Table S1. Specifically, the correction for multiple comparisons was done by constructing an empirical null distribution. For this purpose, $n_p$ = 1000,000 realizations were computed by first randomizing age and then correlating it with all graph metrics at all thresholds and frequency bands, and finally taking maximum correlation value across this permuted correction space. This null distribution is shown in Figure S11. The corrected p-values ($p_c$) were calculated as:

$$p_c = \frac{2(n+1)}{n_p + 1} --- (11)$$

where $n$ is the number values in the empirical null distribution greater or lower than the observed positive or negative correlation value, respectively. The factor of two stems from the fact that the test is two-tailed. Correlations resulting in significant p-values were then again tested using Robust Correlation (Pernet CR et al. 2013), which strictly checks for false positive correlations using bootstrap resampling.

Effect size for correlation was computed using cohen's $d$:

$$d = \left| \frac{2r}{\sqrt{1-r^2}} \right| --- (12)$$

where $r$ is the correlation coefficient.

**2.10. Bootstraping**

For the purposes of visualizing the significance of age effects and assessing uncertainties in the graph metrics with respect to age, we used nested bootstrapping with 1024 realizations.

The nested bootstrap procedure approximates the joint distribution of age $x$ with the age-dependent network metric $f(y_x)$, where $f(y_x)$ is the average network metric over many subjects of age $x$ (see notes below). We observed $n$ pairs $(x_i, y_i)$, where $x_i$ is the age and $y_i$ the corresponding imaging data for the $i^{th}$ subject. Ideally, we would like to observe $(x_i, \bar{y}_x)$, where $\bar{y}_x$ denotes the (conceptual) average of subjects chosen at random from a population, where each subject is of age $x$. Let $f(y)$ denote the function which maps imaging data to a scalar metric describing some aspect of a network. Since $y_i$ contains noise, to visualize and estimate uncertainties in graph metrics we can approximate $(x_i, \bar{y}_x)$ by $(\bar{x}_*, \bar{y}_*)$, where the $*$ denotes a bootstrap sample. We can then evaluate $f(y_{\bar{x}_*})$ instead of $f(y_i)$. Each realization of bootstraping yielded one average network metric and one value for the mean age of the group. Each data point on the normalized density colormap corresponds to one realization of the bootstrap (Figure 2, Figure 4A, B and Figure 5 inserts). Lastly, the correlation values were computed using the original data only, not on the bootsrapped data described here.

### 2.11 LOWESS Regression

We used the non-parametric LOESS regression to fit a curve to the data (Cleveland WS and C Loader). To protect against overfitting in estimating bandwidth, we used 10-fold cross validation. We generated our predictive model using the data in the training set, and then measured the accuracy of the model using the data in the test set. We tested a range of bandwidths from .01 to .99 with .01 step. The bandwidth resulting in least sum of squares error was then selected (Webel K 2006).

**2.12. Machine Learning (Multivariate regression using Random Forest)**

All the graph metrics calculated for each band were combined under a single non-parametric multivariate regression model using Random-Forest (Breiman L 2001; Liaw A and M Wiener 2002). More specifically:

a) Maturity Index (prediction of age using random forest regression model) for beta band (MI-beta) was computed using the beta band network graph metrics: local efficiency, small world property, and resilience.

b) Maturity Index for gamma band (MI-gamma) was computed using the gamma band network graph metrics: global efficiency, small world property and resilience.

c) Maturity Index for both beta and gamma band (MI-combined) was computed using all of the above features.

Lastly, we fit a parametric curve to each of the maturity index, parametric model for curve fitting was selected using Akaike information criterion (AIC). Below we provide details for each step of this procedure.

*2.12.1 Random Forest Regression Analysis*

Random forest (RF), a method for non-parametric regression which is robust in avoiding overfitting (Breiman L 2001), was implemented in R using the randomForest package (Liaw A and M Wiener 2002). As a further step to avoid overfitting, each RF regressor model was trained on the stratified 70% of the data (training set) and then tested in remaining stratified 30% test-set. The possibility of bias introduced by the random choice of the test set was avoided by repeating the sampling 1000 times. The final model represents the aggregate of the 1000 sampling events.

### *2.12.2 Random Forest Parameter optimization*

The optimal number of variables randomly sampled as candidates at each split was selected as p/3 where p is the number of features in the model (Breiman L 2001). A total of 1,000 decision trees were grown to ensure out-of-bag (OOB) error converges (Breiman L 2001).

### *2.12.3 Features (MI-beta, MI-gamma, MI-combined)*

MI-beta (Figure 6A) was derived using machine learning with three features computed in the beta band networks: local efficiency, small world property, and resilience as manifested by the slope of the degree distribution. MI-gamma (Figure 6B) was derived using machine learning with three features computed in the gamma band networks: global efficiency, small world property and resilience as manifested by the slope of the degree distribution. MI-combined (Figure 6C), as its name implies, was computed using machine learning with all six of the above features. The predicted ages for all subjects from the random forest regression model were converted to the maturation indices defined above, using a published scaling scheme (Dosenbach NU *et al.* 2010), by setting the mean predicted brain age to 1, for typically developed young adults. Feature (variable) importance was also computed using random forest regression model by estimating Percent Increase Mean Square Error (%IncMSE). %IncMSE was obtained by permuting the values of each features of the test set and comparing the prediction with the unpermuted test set prediction of the feature (normalized by the standard error)., where %IncMSE is the average increase in squared residuals of the test set when the feature is permuted. A higher %IncMSE value represents a higher variable importance (Figure 6D).

### *2.12.4 Curve Fitting*

The models (Linear, exponential, quadratic, Von Bertalanffy) for best fitted curve (Matlab's Curve Fitting Toolbox) were compared using Akaike information criterion (AIC). Given a set of models for the data, AIC is a measure that assesses the quality of each model, relative to the remaining models in the set. The chosen model minimizes the Kullback-Leibler distance between the model and the ground truth. The model with the lowest AIC is considered the best model among all models specified for the data at hand. The absolute AIC values are not particularly meaningful since they are specific to the data set being modeled. The relative AIC value ($\Delta AIC_i = AIC_i - \min[AIC_p]$) is used to rank models (Akaike H 1974). The model with the minimum AIC was selected as the best model. To quantify goodness of fit, we also computed R-squared (R2; coefficient of determination) for best fitted regression models.

### 2.12.5 Independent Data Set Verification

To verify whether the regression models obtained here were consistent with data collected independently, in a different site, using a different MEG machine (CTF 275), MEG resting state scans from 31 participants from OMEGA project (Niso G *et al.* 2015) were also examined. Random forest regression models learned using our primary, MGH based, dataset, were then applied to the previously unseen OMEGA project data, with no additional training. The predicted maturity indices of the OMEGA sourced subject are shown as orange dots in Figure 6. To quantify goodness of fit on this independent dataset, we also computed R2 between the values from the OMEGA dataset and predictions from models learned from the MGH dataset.

### 2.12.6 Prediction Interval Calculation

The prediction interval for the best fitted curve in Figure 6 was obtained using Scheffe's method (Maxwell SE and HD Delaney 1990).

In this method, the prediction interval $s(\alpha, g)$ is defined as:

$$s(\alpha, g) = \sqrt{g\, F(1 - \alpha, g, n - 2)} ---(11),$$

where g is the model order (2), n is the number of subjects (131), $\alpha$ is the significance level chosen as $0.05/n_c$ ($n_c = 3$; total number of curves), and $F(1 - \alpha, g, n - 2)$ is the F-distribution.

**2.13 Graphical Representation**

Custom code was written in MATLAB to build circle plots for graphically representing connectivity in the three age groups. To represent nodes on the brain, custom code in PySurfer was written using Python. The increases and decreases in connections and topography were represented by using connectivity patterns shown as graphs, with the tool Gephi. This method uses a spring embedding data-driven technique to align regions in two dimensional space based on strength of connections (Fair DA et al. 2008). For Figure 3B, 3D, 4B and 4D, original adjacency matrices were averaged within age-groups and then thresholded for visualizations

## 3. RESULTS

### 3.1 Age-dependent trajectories of network integration by frequency band

The local and global efficiency graph theory metrics were used to evaluate the age-dependent trajectory of local and global network integration, respectively. These metrics were evaluated in each of the five frequency bands. Significant age-dependent changes in local, but not global, efficiency emerged only in the beta band (Figure 2A). In parallel, significant age-dependent changes in global, but not local, efficiency emerged only in the gamma band (Figure 2B). These changes were significant across multiple thresholds (Table S1-A,B). No other significant age dependent changes emerged in any of the other frequency bands (Figure S12). We also tested the weighted network analogues for the same metrics, and the results followed a similar trend (Figure S10). Therefore, the remaining computations were carried out in unweighted networks.

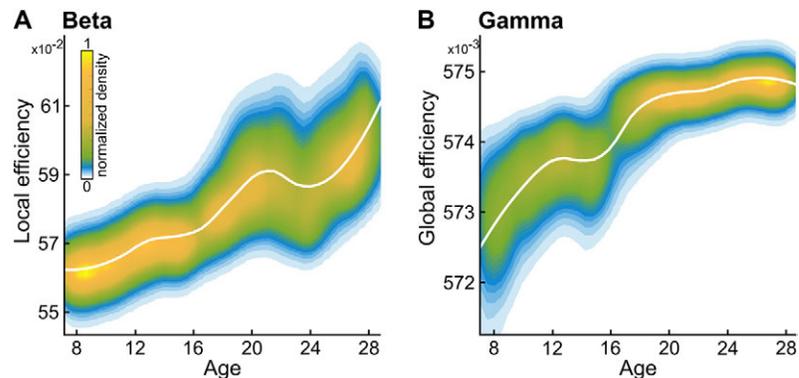

**Figure 2: Network efficiency increases locally in beta band networks and globally in gamma band networks. A.** LOESS plot (solid white line) for the relationship between age and local network efficiency of beta band mediated networks. The individual data points are represented using a normalized density colormap, where each data point corresponds to one realization of the bootstrap procedure. **B.** Same, for gamma band mediated networks, and global efficiency.

### 3.2 Age-dependent trajectories of cortical hubs by frequency band

To assess age dependent changes in spatial distribution of hubs, we measured correlation between age and the betweenness centrality of nodes. In networks mediated by the beta band, loss of betweenness centrality score with age was seen mostly in frontal and temporal hubs, while gain of betweenness centrality score with age was seen mostly in parietal hubs (Figure 3A,B). In networks mediated by the gamma bands, loss of betweenness centrality score with age was seen mostly in occipital hubs, while gain of betweenness centrality score with age was seen mostly in frontal and parietal hubs (Figure 3C,D).

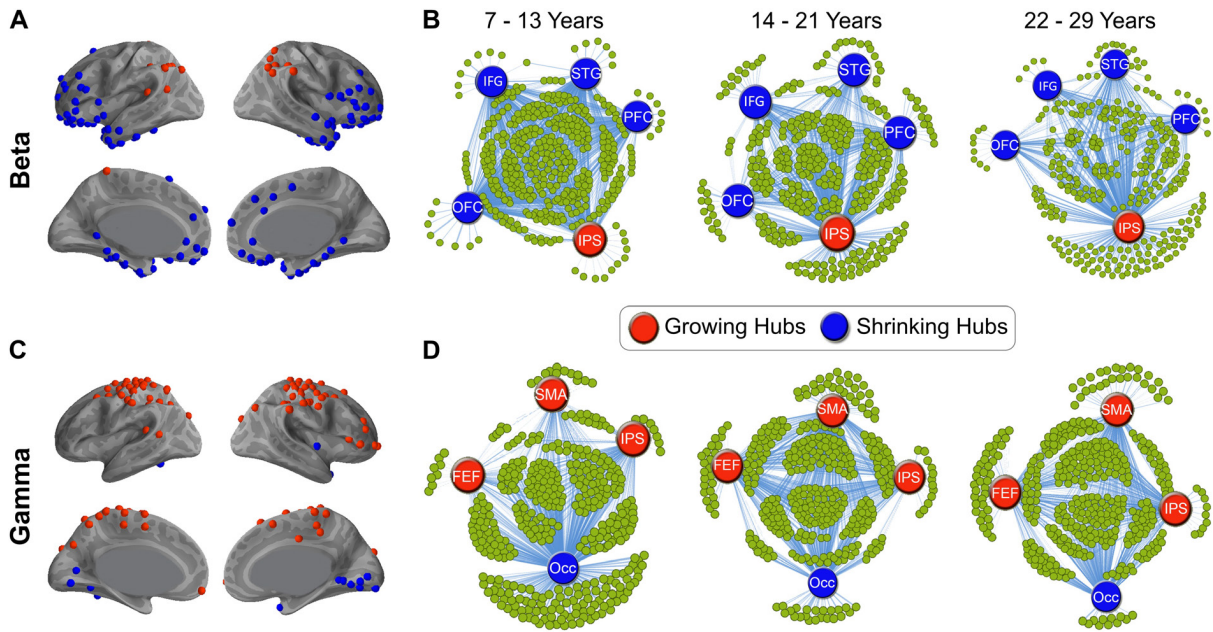

**Figure 3: Spatial distribution and connectivity patterns of growing and shrinking betwnness centrality of hubs.** **A.** Hub regions with growing (red) and shrinking (blue) betweenness centrality scores, in beta band mediated networks. **B.** Visual representation of the connections from 4 hubs with shrinking betweenness centrality scores (blue) and 1 hub with growing betweenness centrality scores (red), averaged for children (7-13), adolescents (14-21) and young adults (22-29), displayed at 0.25 thesholding in beta band mediated networks. **C.** Same as A, for the gamma band mediated networks. **D.** Same as B, for 3 growing and 1 shrinking hubs, for the gamma band mediated networks. **Notation: IPS:** right intraparietal sulcus, **PFC:** right dorsolateral prefrontal cortex, **IFG:** right inferior frontal gyrus, **OFC:** right orbitofrontal cortex, **STG:** superior temporal gyrus, **FEF:** right frontal eye field, **IPS:** right intraparietal sulcus, **SMA:** supplementary motor areas, **Occ:** Occipital cortex.

### 3.3 Age-dependent trajectories of small world property by frequency band

Given the differentiation in age-dependent trajectories between the beta and gamma bands, we next examined the small world property of the networks mediated by each of these frequency bands. While network coefficients in both the beta and gamma bands met small world criteria, we found a significant increase with age in small world property with maturation in the beta band (Figure 4A,B), alongside a significant decrease with age in small world property in the gamma band (Figure 4C,D).

These changes in beta and gamma bands were also consistent across multiple thresholds (Table S1-C for threshold specific p-values).

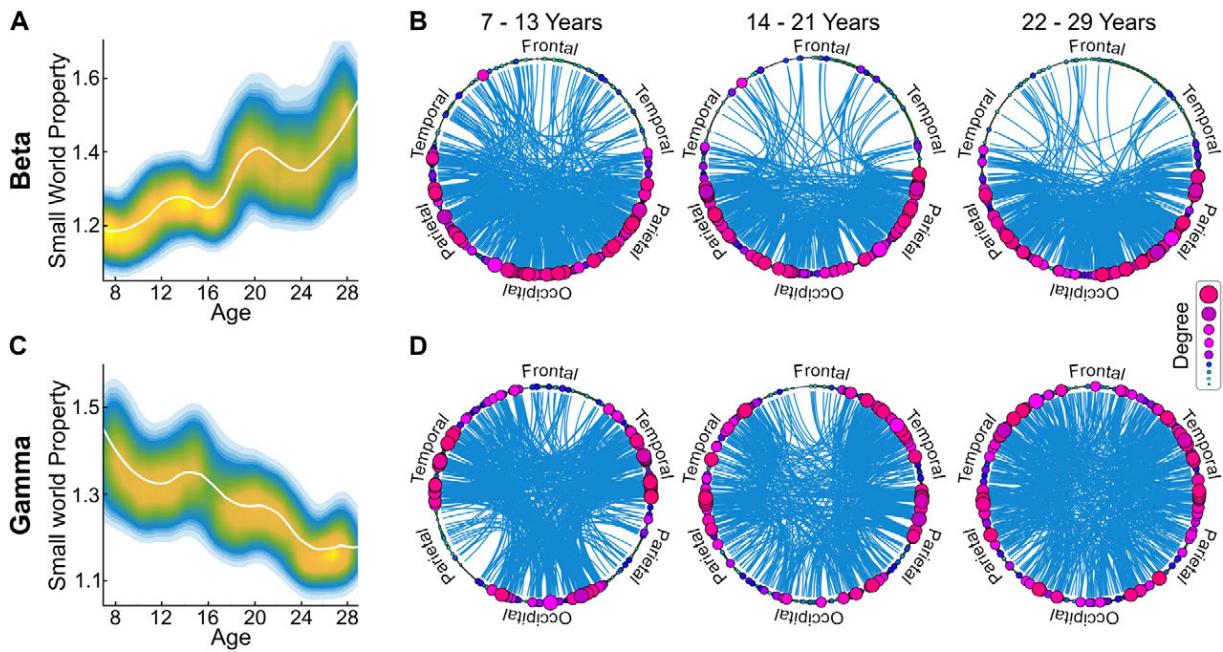

**Figure 4: Small world property is age and frequency band dependent. A.** LOESS plot (solid white line) for the relationship between age and small world property for beta band mediated networks. The individual data points are represented using a normalized density colormap, where each data point corresponds to one realization of the bootstrap procedure (same colorbar as in Figure 2A). **B.** A visual representation of connections and hubs in the three age groups, averaged for children (7-13), adolescents (14-21) and young adults (22-29), in the beta band mediated networks, displayed at 0.25 thesholding. Degree represents the size of the hub. **C.** Same as A, but for the gamma band mediated networks. **D.** Same as B, but for the gamma band mediated networks.

### 3.4 Age-dependent trajectories of network resilience by frequency band

We next evaluated how network resilience, a graph theoretical metric which measures the vulnerability of the network to attacks (by removal) on the most connected hubs, changed with age in beta and gamma band mediated networks. To assess resilience, we quantified the reduction in global efficiency as hubs were removed in order of connectedness, from largest to smallest. We found significant age dependent differences in network resilience in both the beta band (Figure 5A) and the gamma band (Figure 5B) mediated networks. While the significance of the age effect differed by percentage of nodes removed, whenever there was a significant age effect, the trend was similar; resilience weakened with age in the beta band mediated networks, but strengthened with age in the gamma band mediated networks, as shown in the insets of Figure 5. We repeated the same analysis with local efficiency as the parameter, as well as with attacking the connections rather than the hubs. The results followed the same trajectory in all cases, with resilience weakening with age in the beta band, and increasing with age in the gamma band.

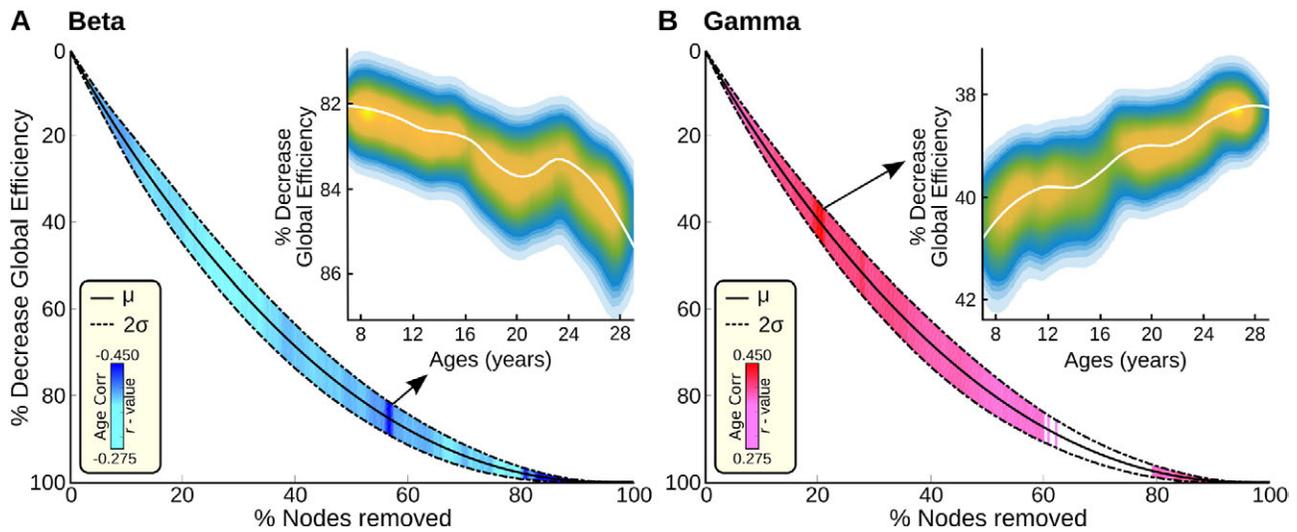

**Figure 5: Resilience in Gamma and beta mediated networks follows opposite developmental trajectories. A:** Main plot - Solid line mean resilience, combined for all ages, plotted as the decrease in global efficiency as a function of percent nodes removed in beta band mediated networks, from largest to smallest. Dashed lines mark confidence interval at two standard deviations. For each point on the average curve, we computed whether there was a significant age effect. When age was a factor in resilience, that line between the upper and lower confidence interval for that point on the curve was assigned a color. The color marks the correlation coefficient of the effect of age, thresholded at p<0.05 corrected (i.e. at significance). The colorbar at the bottom left shows the colormap of strength of the correlation coefficient. **A-Inset:** LOESS plot of one instance of the effect of age, at 54% of nodes removed, where significance of age effect was maximal. The individual data points are represented using a normalized density colormap, where each data point corresponds to one realization of the bootstrap procedure (same colorbar as in Figure 2A). While the exact numbers differed for different percent nodes removed, the pattern was always identical, showing a negative correlation with age for the gamma band mediated networks. **B:** Main plot - Same as A, but for gamma band mediated networks. The white patch between 60% and 80% nodes removed indicates there was no impact of age in that range. **B-Inset:** Same as A-Inset, for the beta band, at 19% of nodes removed, where significance of age effect was maximal. Again, whenever there was a significant age effect, marked by colors on the main curve, the pattern was identical, showing a positive correlation with age for the beta band mediated networks.

**3.5 Age prediction by frequency band specific properties**

We next tested whether the graph metrics assessed here could be used to predict individual brain maturity. Given the different trajectory observed in the beta and gamma band mediated networks, we began by assessing age-based prediction using random forest regression within each set of networks separately. For the beta band, we defined a beta Maturity Index, MI-beta, using local efficiency, small world property, and resilience parameters from the beta mediated networks. When MI-beta was plotted relative to age, we found that age prediction follows a linear trajectory (Figure 6A). For the gamma band, likewise, we defined a gamma maturity index, MI-gamma, using global efficiency, small world property, and resilience parameters from the gamma mediated networks. When MI-gamma was plotted relative to age, we found that prediction followed a non-linear quadratic asymptotic growth curve trajectory (Figure 6B). Combining the two sets of measures resulted in a non-linear Von Bertalanffy growth curve that yielded significantly increased prediction accuracy (Figure 6C). The relative information (explained variance) of each of the parameters is shown in Figure 6D. MI-beta and MI-gamma together accounted for 52% of the variance observed in the data. To further test the reliability of the model, which was trained solely on MGH data set, it was applied blindly to 31 participants from an independent dataset (OMEGA), with no additional training. The data were plotted alongside the MGH data, and indeed follow the same trajectories (Figure 6A-C).

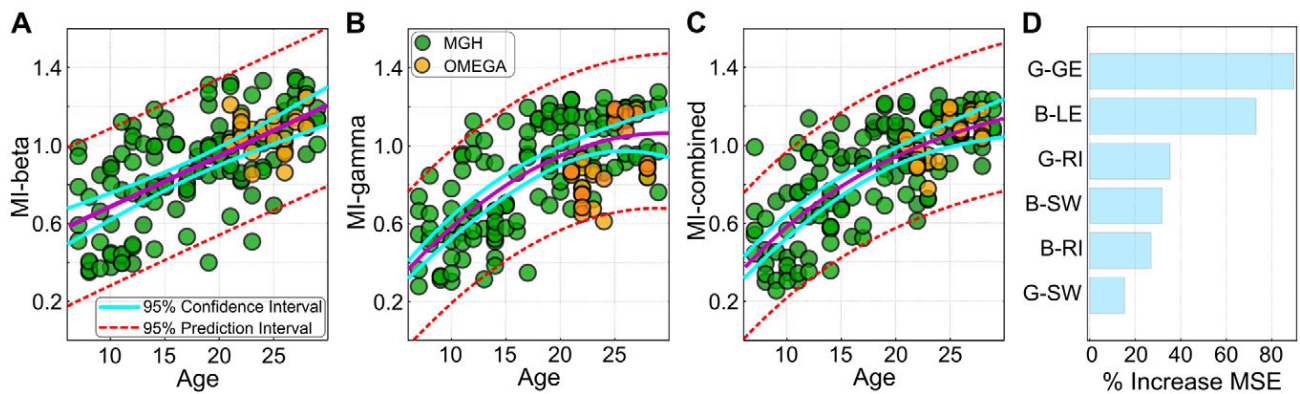

**Figure 6: Classification by maturation curve.** Green circles represent results for individual study participants ("MGH") Orange circles represent values for participants from the independent OMEGA database (OMEGA), that were not used during the learning phase in the machine learning analysis. **A.** MI-beta ($R^2$=0.39 MGH, $R^2$ = 0.34 OMEGA) plotted relative to age. **B.** MI-gamma ($R^2$=0.48 MGH, $R^2$=0.33 OMEGA) plotted relative to age. **C.** The combined MI for beta and gamma ($R^2$=0.52 MGH, $R^2$=0.41 OMEGA) plotted relative to age. **D.** The relative contribution (variable importance computed using random forest regression) of each of the parameters to the model. Notation: GE: global efficiency; LE: local efficiency; RI: Resilience index (see SM). SW: Small world property. Orange circles in panels A-C mark the participants from the OMEGA data set.

## 4. DISCUSSION

We found that from age 7 to 29, resting state networks mediated by the beta and gamma frequency bands underwent marked topological reorganization, while resting state networks mediated by the slower alpha, theta and delta frequency bands showed no significant age dependent changes in network topology, for the examined graph-theoretical metrics. Importantly, the patterns of age-dependent changes for the beta and gamma mediated networks differed substantially. Beta band mediated networks became more locally efficient, i.e. tending towards clustering and more connections with adjacent regions with age, while gamma band mediated networks became more globally efficient, i.e. tending towards shorter overall path lengths and thus faster communication across larger cortical distances, with age. Additionally, the contribution and importance of many hubs to the overall network efficiency, measured using betweenness centrality, grew or shrunk with age, but a different set of hubs showed this pattern for beta and gamma mediated networks, with relatively little overlap. Since small world property and resilience are inversely proportional to one another and both depend on the relative magnitude of local and global efficiencies, these measures presented opposite age-dependent trajectories for the beta and gamma mediated networks. Specifically, small world property, i.e. overall network optimization in balancing short and long connections, increased with age, while resilience, i.e. robustness of the network, decreased with age in the beta band. The pattern was exactly opposite in the gamma band mediated networks. Remarkably, the two sets of networks followed different growth trajectories, with the beta band mediated networks best described with a linear, rather than an asymptotic, growth curve, and the gamma band mediated networks best described by a more expected asymptotic growth curve (Dosenbach NU *et al.* 2010).

These results extend prior fMRI based findings in several ways, most notably by determining that only two out of the five fundamental frequency bands, beta and gamma, mediated the resting state networks that showed age-related changes, and that each followed a distinct trajectory, and in the case of the beta band, that trajectory was unexpectedly linear. Furthermore, contrary to prior suggestions from fMRI based studies (Fair DA *et al.* 2009; Hwang K et al. 2013), we found that the small world property, which assesses the overall balance of the network in optimizing local versus distant connections, did not remain constant through this age range, and similarly, network resilience at a given age depended on the underlying frequency. Lastly, as reported with fMRI (Fransson P et al. 2011; Menon V 2013), gamma band mediated networks showed development of hubs in heteromodal regions such as posterior parietal, posterior cingulate and the anterior insula. But unlike observations in fMRI studies, beta band mediated networks showed a loss of hubs in heteromodal-frontal regions, alongside growth in hubs in posterior parietal regions.

The observation that only the resting state networks that were mediated by the gamma and beta frequency bands showed significant topological reorganization with age may be driven by the fact that these two bands are strongly associated with cognitive control (Buschman TJ and EK Miller 2014; Roux F and PJ Uhlhaas 2014), which matures over adolescence (Luna B et al. 2015). It is likely also related to the fact that both of these high frequency rhythms are heavily dependent on GABAergic systems (Uhlhaas PJ *et al.* 2008; Sohal VS et al. 2009), which themselves undergo extensive changes during development, well into adolescence. The pattern of reduced frontal hubs observed in the beta band is in line with observations showing reduced frontal task related activation with maturation, for instance for inhibitory control, potentially due to increased efficiency of top-down communication, putatively mediated by the beta band (Ordaz SJ et al. 2013). In particular, the linear growth trajectory of the beta band mediated networks could be the result of the continuing maturation and development of top-down projections, which may be more likely to be mediated via the beta band (Buschman TJ and EK Miller 2007; Wang XJ 2010). Indeed, many processes that are heavily mediated via top-down connections, such as attention and verbal functioning, peak past the age range examined here (Peters BD et al. 2014). In contrast, the gamma band mediated networks followed the more expected asymptotic trajectory, which may reflect the completion of maturation of bottom-up projections, which have been associated with greater probability with the gamma band (Buschman TJ and EK Miller 2007; Wang XJ 2010).

The results using the resilience metric, i.e the measure of the robustness of networks, are particularly intriguing in the context of psychiatric disorders (Lo CY *et al.* 2015). In our prior studies, we have found that resting state networks in ASD, ages 8-18, showed increased efficiency in the gamma band, but decreased efficiency in the beta band (Kitzbichler MG *et al.* 2015). In parallel, studies have shown reduced efficiency in the alpha band in bipolar disorder and schizophrenia (Hinkley LB et al. 2011; Kim DJ et al. 2013; Kim JS et al. 2014), and abnormal resting state network connectivity in the gamma band (Andreou C et al. 2015). The observation of resilience, a measure that depends on network efficiency, increased with age in the beta band but decreased with age in the gamma band, diverges from our original hypothesis of minimal resilience during adolescence. However, it is possible that greater vulnerability during adolescence arises from the fact that resilience is not optimized in this age range in either network, and thus both are relatively more vulnerable.

The study does have several limitations the merit noting. One limitation is that we chose to focus on eyes open with relaxed fixation as our resting state paradigm, rather than eyes closed, thus minimizing alpha power. This was done to best align with parallel prior fMRI studies (Fair DA et al. 2007; Fair DA *et al.* 2009; Dosenbach NU *et al.* 2010; Grayson DS et al. 2014), and to follow the guidelines of the Human Connectome Project. Eyes-open resting state networks derived using MEG also have greater test-retest reliability than eyes-closed derived networks (Jin SH et al. 2011). While some MEG/EEG studies do find differences between the two conditions (Jin SH *et al.* 2011; Tan B et al. 2013; Tagliazucchi E and H Laufs 2014; Miraglia F et al. 2016; Yu Q et al. 2016), overall, the differences between the eyes closed and eyes open conditions in all of these studies were small. Another limitation of the proposed study is that we only had IQ measures available for a subset of the sample (N=68), and no other behavioral measures uniformly across the sample. While we were able to show that there is no relationship to IQ in the subset of the sample for which IQ was available (Figure S2), the absence of behavioral assessments means we were not able to link the measures to any specific cognitive measures.  A minor limitation is a different in head size across development. Given that brain size reaches 95% of its maximum size by age 6 (Giedd JN et al. 1999; Lenroot RK and JN Giedd 2006), and our minimum age is 7, the impact of changing brain size is likely slight (see also Figure S3 and methods 2.5.2), but cannot be completely dismissed. Another important limitation is that this study focuses solely on topological network properties. Developmental studies of coherence for instance, clearly show increased coherence in the beta band as well in the alpha band (Schäfer CB *et al.* 2014). Indeed, when we look at changes with age of "degree", which measures the mean functional connectivity of each node, we find age dependent changes in the alpha band as well (Figure S9), reproducing these prior results. Further studies will need to be carried out to elucidate the contributions and relevance of topological versus more direct non-topological properties such as coherence, to cognitive development. Lastly, the independent data set only spans ages 21-28, and not the full age range studied here. As of now, unfortunately, there are no pediatric shared data sets of resting state MEG data. Therefore, our independent data set was by necessity limited in age range.

In summary, we show that developmental refinement of resting state networks as assessed by graph metrics is dependent on the mediating frequency band, and age dependent changes in global network properties occur only in the beta and gamma bands between the ages of 7 and 29. Specifically, we show that gamma band mediated networks become more globally efficient with maturation, while beta band mediated networks become more locally efficient with maturation. Accordingly, the small world property, which measures how optimally balanced local and global efficiencies are, increased with age in beta band mediated networks, and decreased with age in gamma mediated networks. To reconcile our results with prior fMRI findings on the development of topological network properties, we need to consider the fact that since fMRI signal cannot be used to distinguish signals from different frequency bands (Hipp JF and M Siegel 2015), prior fMRI-based studies observed results from all frequency bands combined. In such a scenario, it might indeed appear that the small world property remains unchanged with age, as would resilience, if the signals from the beta and gamma band were weighed roughly equally in the fMRI signal. Similarly, because the combination of a linear and a non-linear function results in a non-linear function, as shown in Figure 6C, the linear maturation trajectory observed here in beta mediated networks would likely be missed by fMRI. This observed differentiation between beta and gamma band mediated networks could hint at underlying neural mechanisms in case of abnormal maturation. For instance, disorders that are more impacted in the gamma band might be more related to dysfunction in PV+ interneurons (Takada N et al. 2014). In contrast, disturbances in maturation in the beta band might be more attributable to inhibitory-inhibitory connections (Jensen O et al. 2005). In combination, our findings significantly advance our understanding of the complex dynamics behind oscillatory interactions that subserve the maturation of resting state cortical networks in health, and their disruptions in developmental and psychiatric or neurological disorders.

## 5. ACKNOWLEDGMENTS:

This work was supported by grants from the Nancy Lurie Marks Family Foundation (TK, SK, MGK), Autism Speaks (TK), The Simons Foundation (SFARI 239395, TK), The National Institute of Child Health and Development (R01HD073254, TK), The National Center for Research Resources (P41EB015896, MSH), National Institute for Biomedical Imaging and Bioengineering (5R01EB009048, MSH), and the Cognitive Rhythms Collaborative: A Discovery Network (NFS 1042134, MSH).

# Supplementary Materials:

## Maturation Trajectories of Cortical Resting-State Networks Depend on the Mediating Frequency Band

S. Khan[1,4,5‡*], J. A. Hashmi[1,4‡], F. Mamashli[1,4], K. Michmizos[1,4], M. G. Kitzbichler[1,4], H. Bharadwaj[1,4], Y. Bekhti[1,4], S. Ganesan[1,4], K. A Garel[1,4], S. Whitfield-Gabrieli[5], R. L. Gollub[2,4], J. Kong[2,4], L. M. Vaina[4,6], K. D. Rana[6], S. S. Stufflebeam[3,4], M. S. Hämäläinen[3,4], T. Kenet[1,4]

[1]Department of Neurology, MGH, Harvard Medical School, Boston, USA.
[2]Department of Psychiatry MGH, Harvard Medical School, Boston, USA.
[3]Department of Radiology, MGH, Harvard Medical School, Boston, USA.
[4]Athinoula A. Martinos Center for Biomedical Imaging, MGH/HST, Charlestown, USA
[5]McGovern Institute for Brain Research, Massachusetts Institute of Technology, Cambridge, USA
[6]Department of Biomedical Engineering, Boston University, Boston, USA

‡equal contribution

Running Title: Multifaceted Development of Brain Connectivity with Age

[*]**Corresponding author:**
Sheraz Khan, Ph.D.
Athinoula A. Martinos Center for Biomedical Imaging
Massachusetts General Hospital
Harvard Medical School
Massachusetts Institute of Technology
149 13th Street, CNY-2275
Boston, MA-02129, USA
Phone: +1 617-643-5634
Fax:  +1 617-948-5966
E-mail: sheraz@nmr.mgh.harvard.edu

### A. Local Efficiency (p-values)

| Thresholds (Costs) | 0.05 | 0.1 | 0.15 | 0.2 | 0.25 | 0.3 |
|---|---|---|---|---|---|---|
| Delta | 0.90858 | 0.35435 | 0.34725 | 0.53959 | 0.71032 | 0.75228 |
| Theta | 0.31430 | 0.53240 | 0.59036 | 0.26965 | 0.13744 | 0.11737 |
| Alpha | 1.00000 | 0.80481 | 0.54464 | 0.42790 | 0.31150 | 0.24364 |
| Beta | 0.00002 / 1.1413 | 0.00001 / 1.1683 | 0.00004 / 1.1065 | 0.00005 / 1.0914 | 0.00006 / 1.0850 | 0.00005 / 1.0902 |
| Gamma | 0.12228 | 0.83435 | 0.34341 | 0.07796 | 0.03301 / 0.6158 | 0.02289 / 0.6541 |

### B. Global Efficiency (p-values)

| Thresholds (Costs) | 0.05 | 0.1 | 0.15 | 0.2 | 0.25 | 0.3 |
|---|---|---|---|---|---|---|
| Delta | 0.99998 | 0.99534 | 0.99342 | 0.99995 | 0.99985 | 0.99917 |
| Theta | 0.89226 | 0.93476 | 0.65848 | 0.64691 | 0.47928 | 0.98805 |
| Alpha | 0.98163 | 0.78137 | 0.68508 | 0.54658 | 0.60303 | 0.56566 |
| Beta | 0.55083 | 0.10113 | 0.05066 | 0.07189 | 0.11197 | 0.47393 |
| Gamma | 0.00122 / 0.8943 | 0.00028 / 0.9905 | 0.00025 / 0.9987 | 0.00025 / 0.9987 | 0.00064 / 0.9380 | 0.20087 |

### C. Small World Property (p-values)

| Thresholds (Costs) | 0.05 | 0.1 | 0.15 | 0.2 | 0.25 | 0.3 |
|---|---|---|---|---|---|---|
| Delta | 0.99965 | 1.00000 | 1.00000 | 1.00000 | 0.99961 | 0.99604 |
| Theta | 0.47492 | 0.61659 | 0.56560 | 0.29959 | 0.27750 | 0.31892 |
| Alpha | 0.99979 | 0.99942 | 0.90882 | 0.75562 | 0.56541 | 0.45317 |
| Beta | 0.00216 / 0.8541 | 0.00020 / 1.0116 | 0.00010 / 1.0537 | 0.00014 / 1.0306 | 0.00003 / 1.1213 | 0.00004 / 1.1095 |
| Gamma | 0.13045 | 0.03978 / 0.5951 | 0.02702 / 0.6372 | 0.01505 / 0.6949 | 0.02123 / 0.6617 | 0.01090 / 0.7243 |

**Table S1: Corrected (see methods) correlation p-values over the range of thresholds used, for each frequency band**. Colormap codes for significant p-values. Cohen's *d* for correlations effect sizes are shown in blue. **(A)** P-values for local efficiency, per band, per threshold value. The p-value in the blue rectangle corresponds to Figure 2A. **(B)** P-values for global efficiency, per band, per threshold value. The p-value in the blue rectangle corresponds to Figure 2B. **(C)** P-values for small world property, per band, per threshold value. The p-values in the blue rectangles correspond to Figure 4A (beta band) and 4C (gamma band). **All graph derived figures throughout the manuscript and SI are shown at 15% thresholding (cost)**. P-Values were corrected for multiple comparisons (Frequency bands, thresholds, graph metrics) by controlling for family-wise error rate using maximum statistics through permutation testing (see also section "Correction of correlation p-values for multiple comparisons", and Figure S8 for the null distribution).

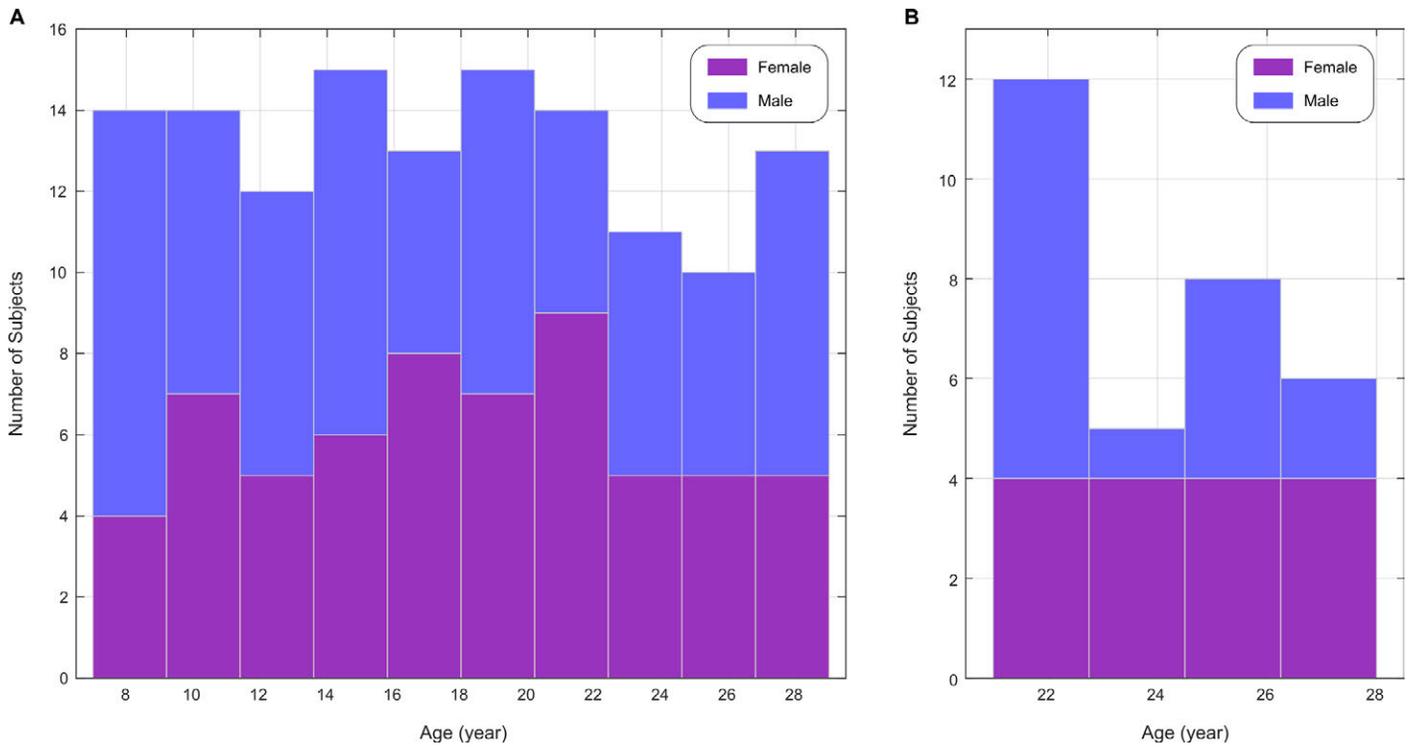

**Figure S1:** Subject distribution for (A) our data (MGH), and (B) the OMEGA data

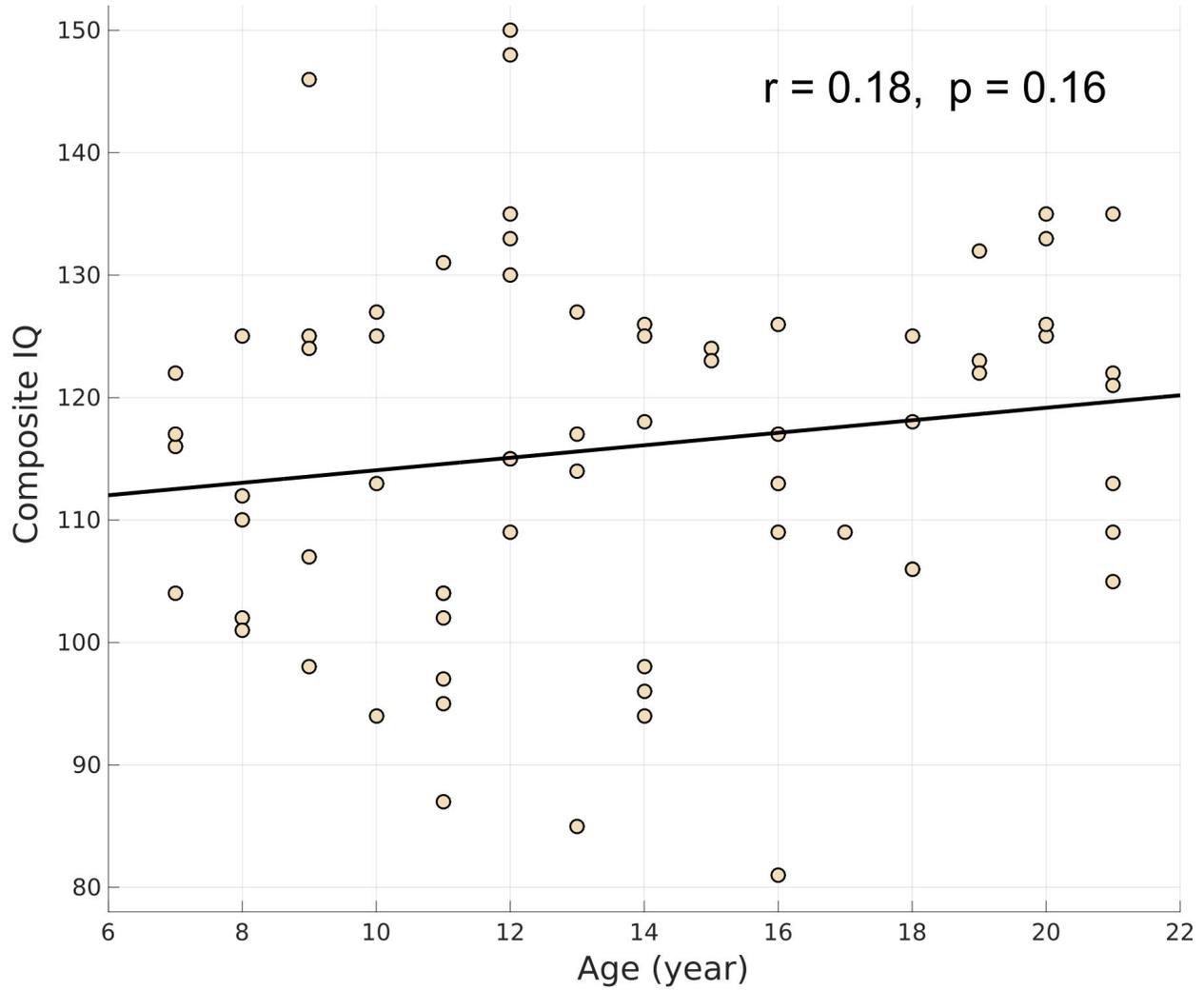

**Figure S2: Correlation between age and Composite IQ (r = 0.18, p=0.16).** Linear regression line (solid black line) for the relationship between age and IQ. The individual data points are represented using a scatter plot. There was no significant relationship between age and IQ, as expected.

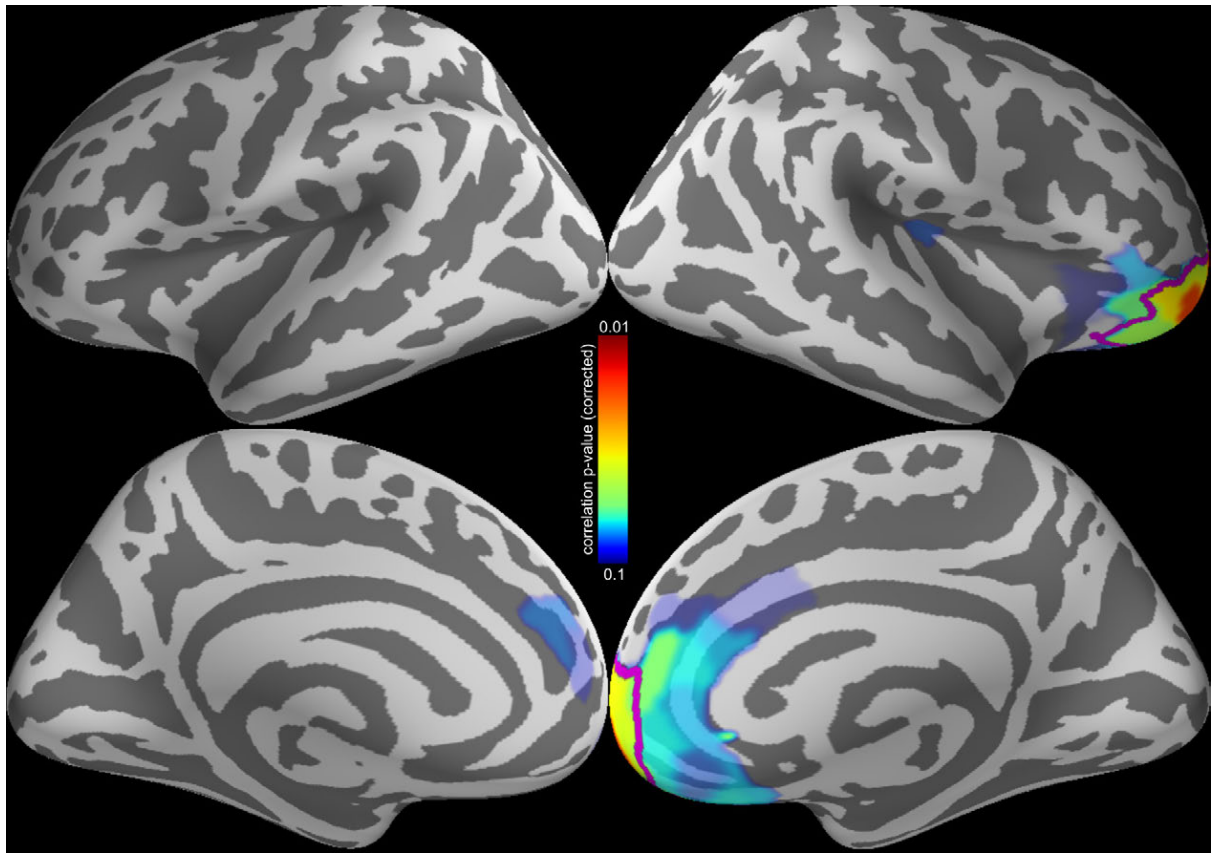

**Figure S3: Correlation between age and norm of the lead field for gradiometers.**
Thresholded p-value for this correlation are shown as textured colormap on the cortical manifold. Magenta line outlines the only areas with significant correlation, at p < 0.05. These areas did not overlap with any of the identified hubs in the study, meaning age effects due to head size were likely negligible in the presented analyses. Note that no significant correlation with age was observed in the magnetometer data.

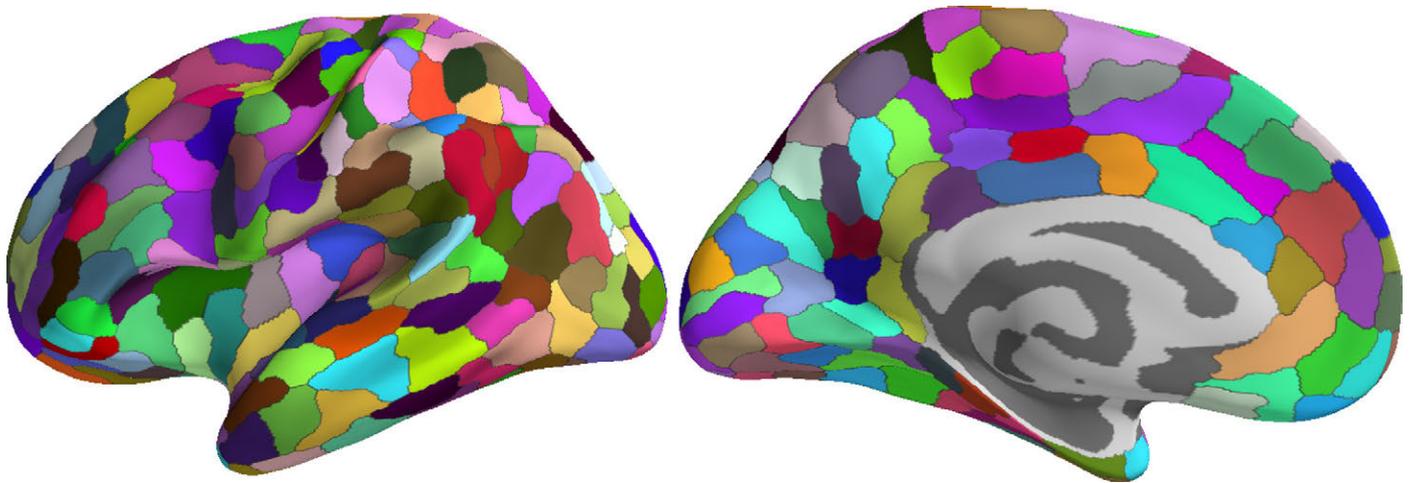

**Figure S4:** Parcellation Scheme on the cortical surface showing 448 labels.

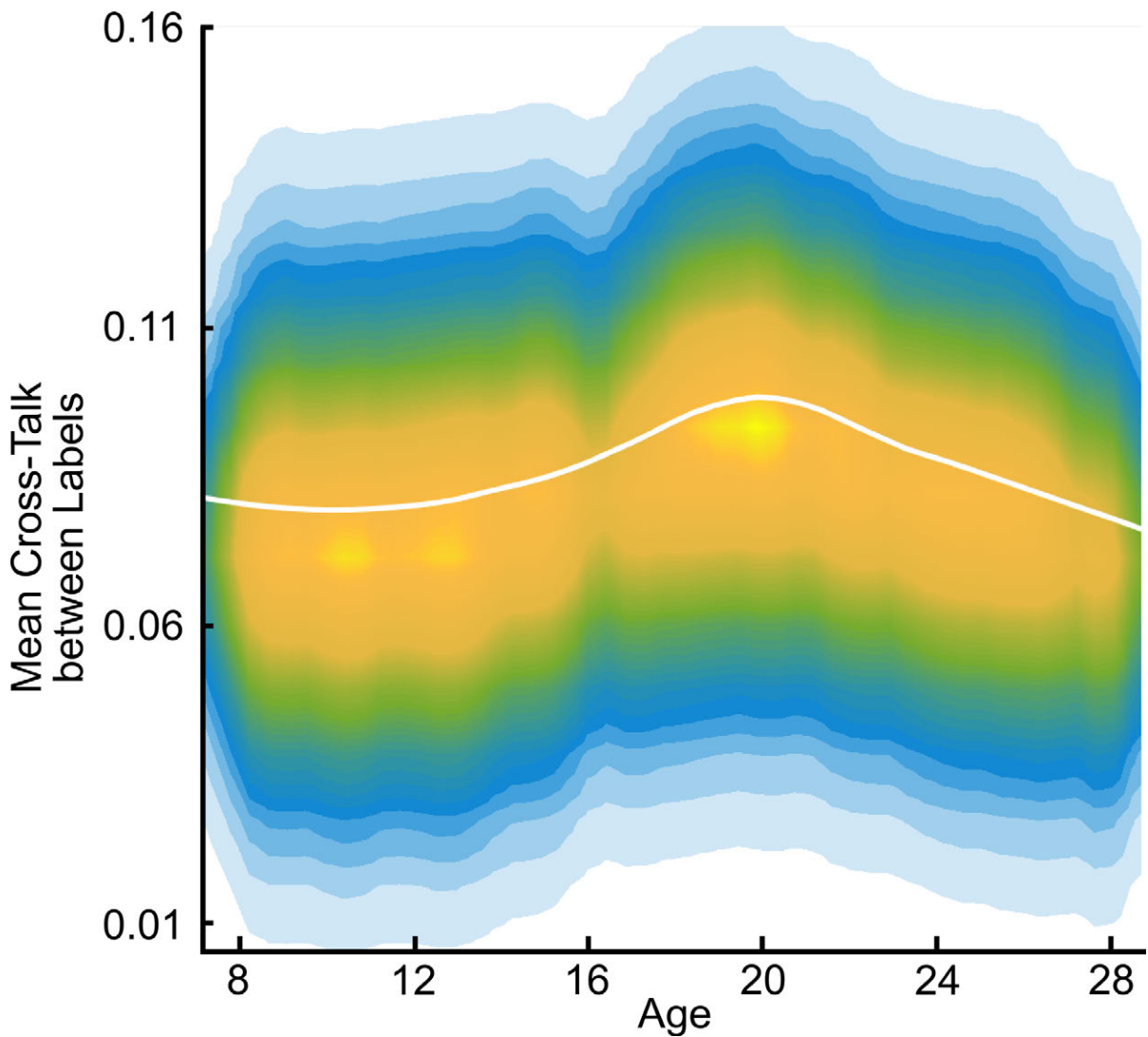

**Figure S5: Correlation between age and mean cross-talk for 448 Labels (p = 0.36). A**. LOESS plot (solid white line) for the relationship between age and mean cross-talk between labels. The individual data points are represented using a normalized density colormap, where each data point corresponds to one realization of the bootstrap procedure.

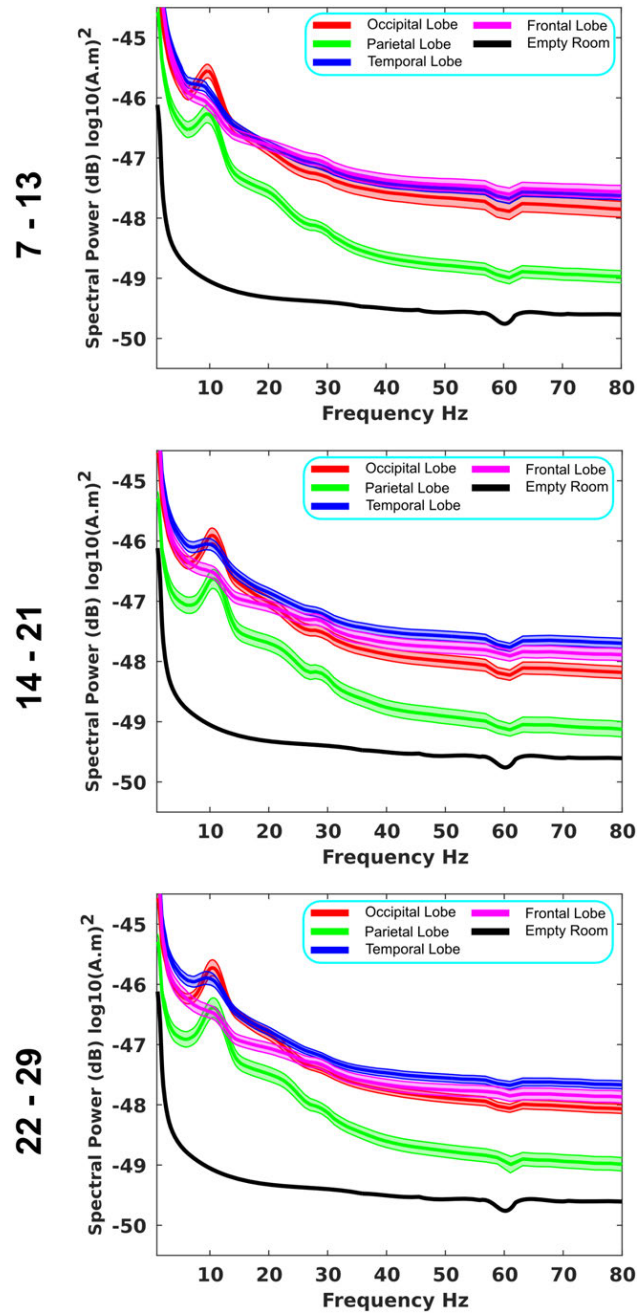

**Figure S6: Power Spectral Density in different cortical areas (source space data).** The PSD plots are averaged across all participants in each of the three age groups noted above. Legend shows cortical area from which data was derived (source space). The lines show the mean PSD within that age group and cortical region, averaged over the entire region. Standard error is shown as shaded area. Empty room data mapped onto the cortical space is shown as comparison for estimated SNR. (**A**) Ages 7 – 13. (B) Ages 14 – 21.  (**C**) Ages 22 – 29. Red: Occipital Lobe; Green: Parietal Lobe; Blue: Temporal Lobe; Blue: Temporal Lobe; Magenta: Frontal Lobe, Empty Room: Mapped onto the cortical Manifold.

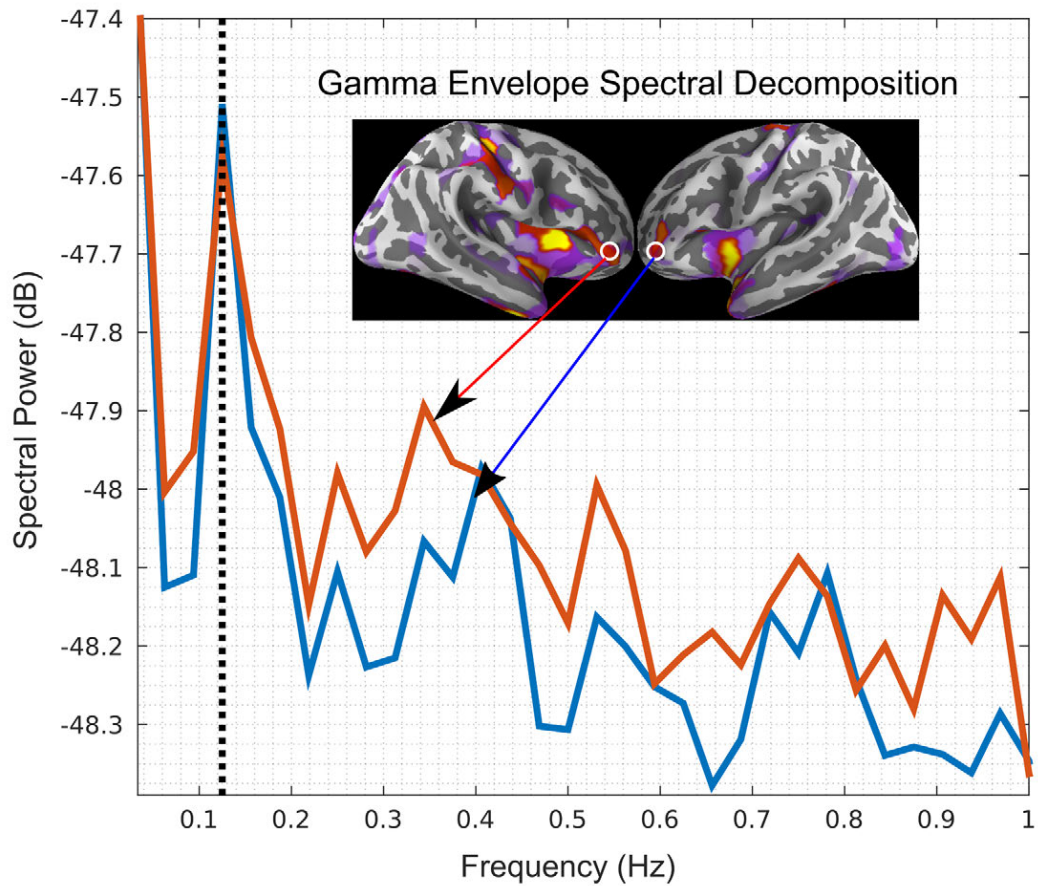

**Figure S7: Example of the envelope approach applied to gamma frequency band (31Hz-80Hz)**: Envelopes of gamma power in two cortical regions (left and right lateral orbitofrontal cortex) in one participant, marked in red (right) and blue (left). The peak frequency of the two independently derived envelopes is around 0.125Hz. The correlation between two regions is then measured via the correlation in these envelopes, not the carrier frequency bands. This is also illustrated in the fourth panel (fourth step) in Figure 1.

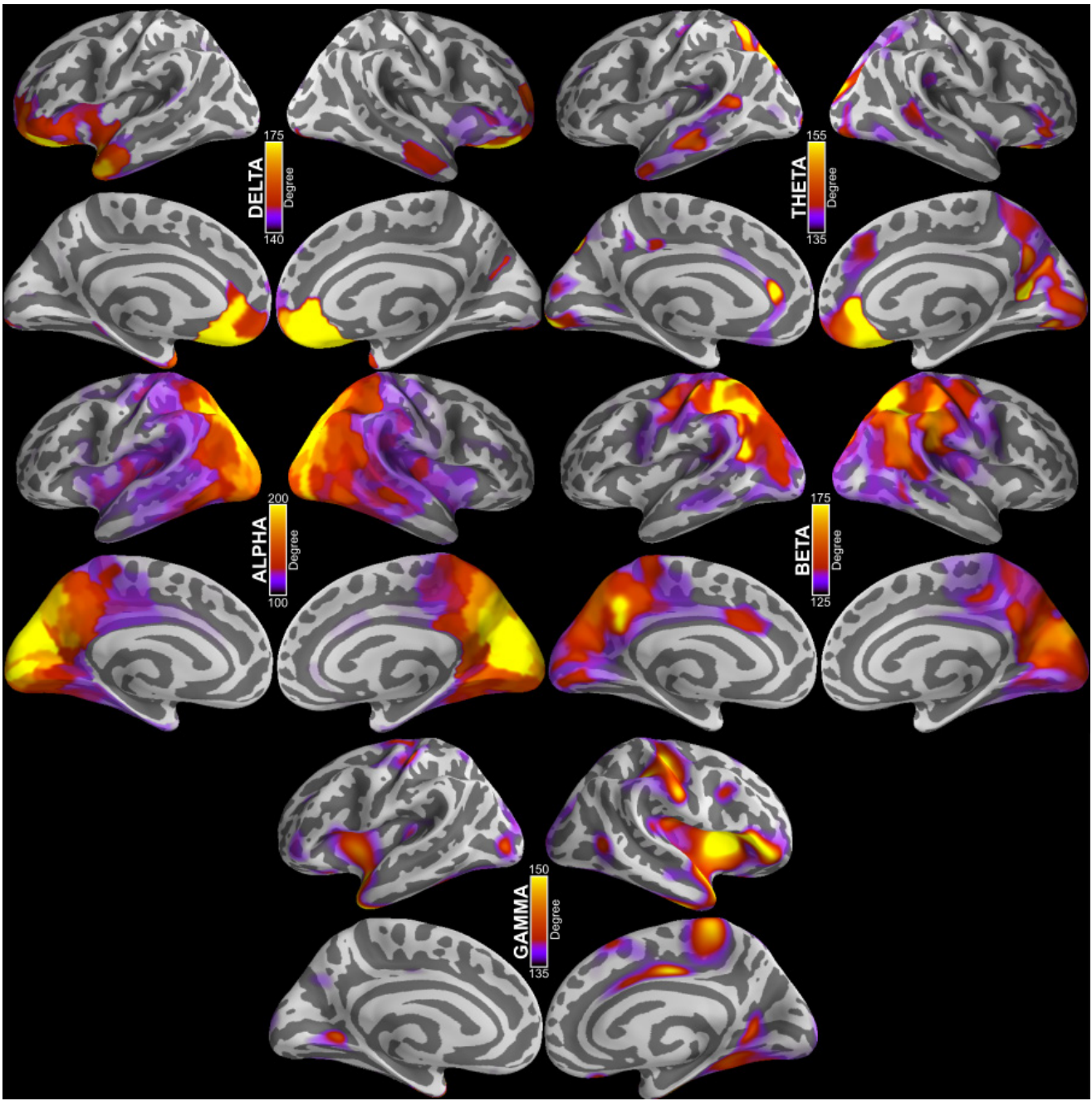

**Figure S8: Degree of each node for age group 22-29, for each frequency band.** The textured colormap on cortical manifold shows mean degree across subjects in age group 22-29 for each label. There maps look similar to prior studies (Hipp et al. 2012). Note that the original Hipp et al study was done in cortical volume space and then projected onto the surface, whereas our study directly computes all the results on the surface. Both approaches show sensible networks. Please see methods for additional details.

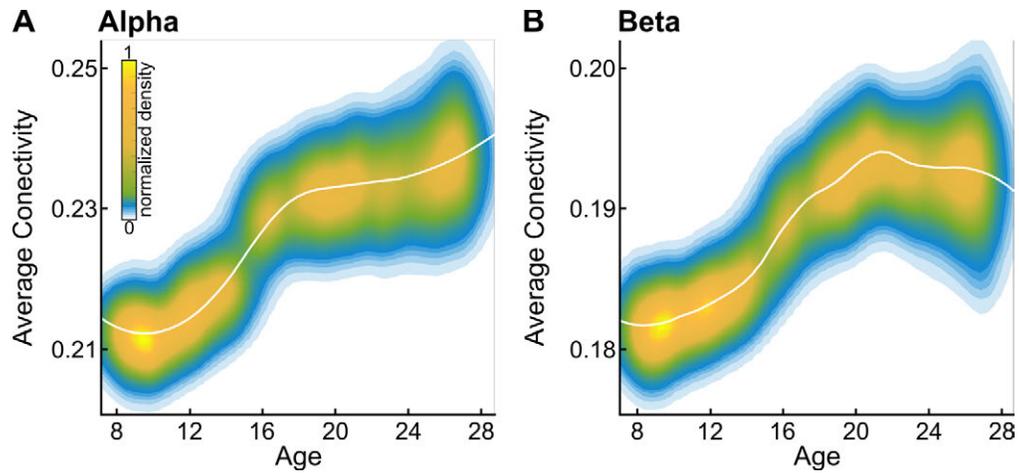

**Figure S9: Average connectivity increases with age in the alpha and beta band mediated networks.** A. LOESS plot (solid white line) for the relationship between mean connectivity (unthresholded weighted degree, also known as node strength) and age for alpha band mediated networks. The individual data points are represented using a normalized density colormap, where each data point corresponds to one realization of the bootstrap procedure. B. Same, for the beta band mediated networks.

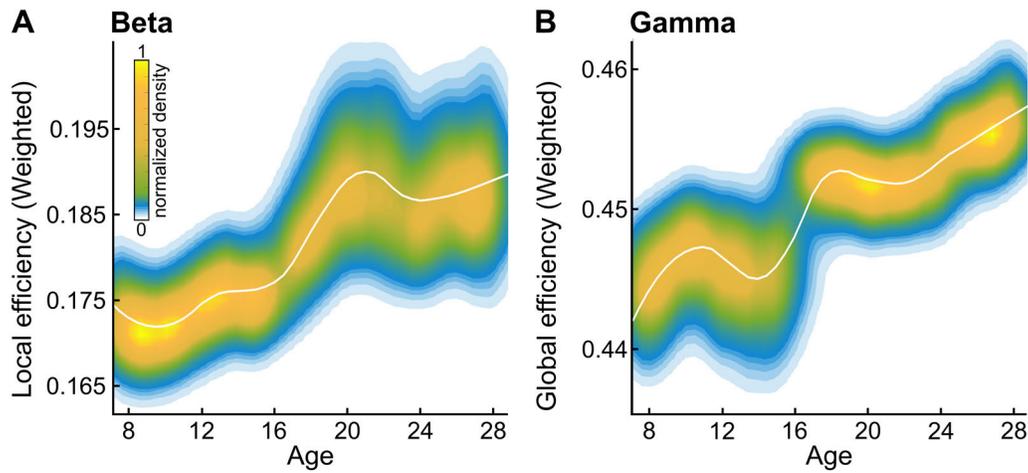

**Figure S10: Weighted Networks - local efficiency increases in beta band networks and global efficiency increases in gamma band networks.** **A**. LOESS plot (solid white line) for the relationship between age and weighted local network efficiency of beta band mediated networks. The individual data points are represented using a normalized density colormap, where each data point corresponds to one realization of the bootstrap procedure. **B**. Same, for gamma band mediated networks, and weighted global efficiency. The results show the same trend as unweighted networks, shown in Figure 2.

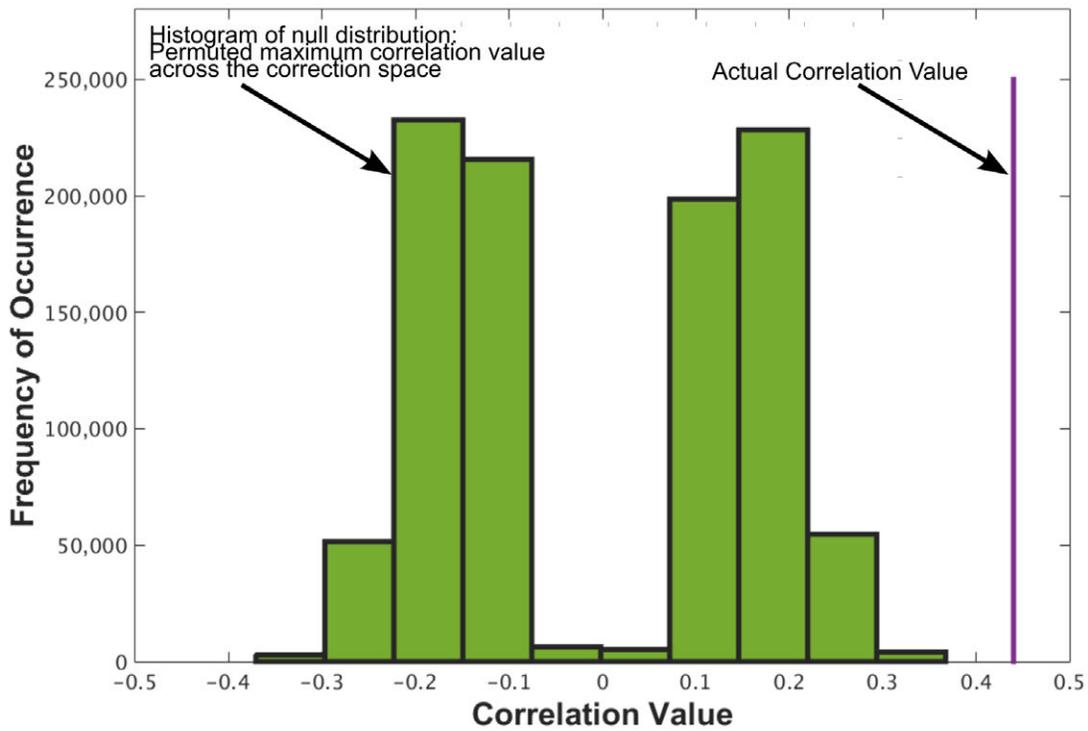

**Figure S11: Calculating Corrected Correlation p-value.** The null distribution (1000,000 Permutations) was computed as described in the method section, and is shown. For each actual correlation value in the data shown as magenta line, we compute the corrected p-value by comparing it against the empirical null distribution.

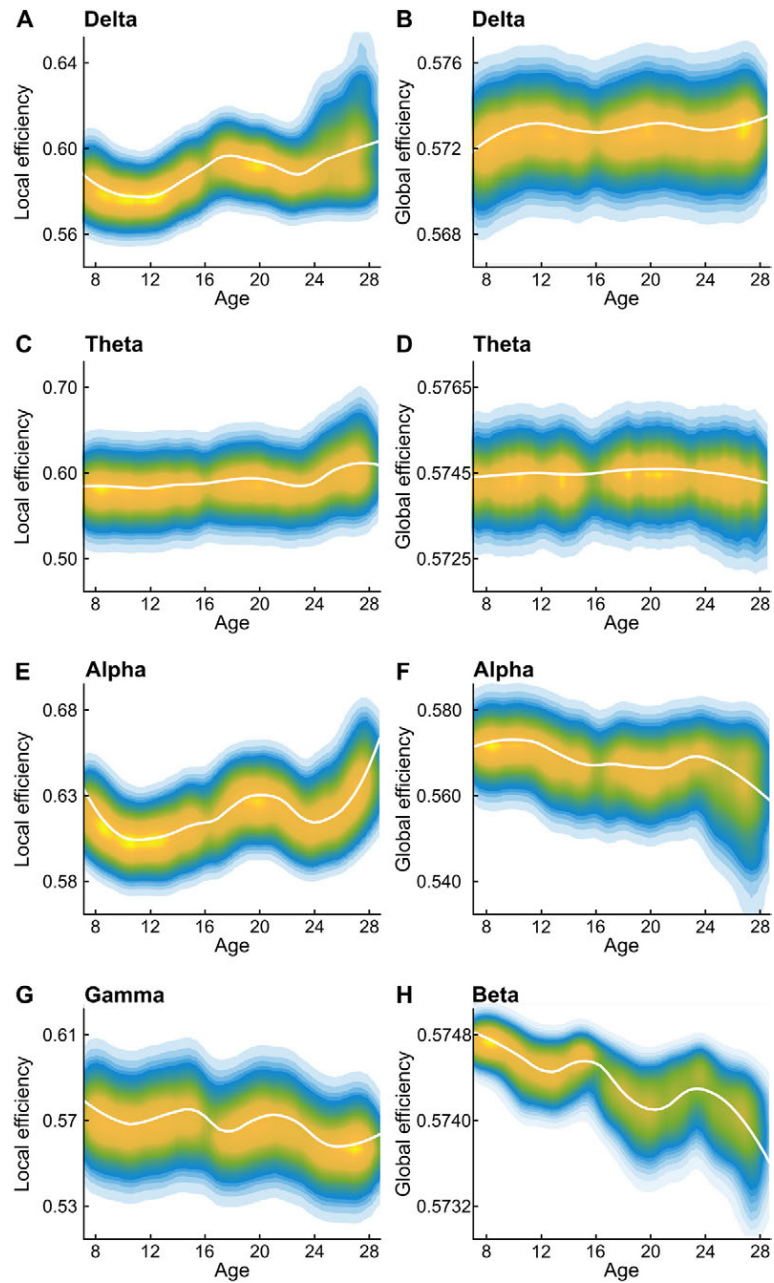

**Figure S12: Local and global network efficiencies did not show age dependent changes in the remaining cases**. **A.** LOESS plot (solid white line) for the relationship between age and local network efficiency of Delta band mediated networks. The individual data points are represented using a normalized density colormap, where each data point corresponds to one realization of the bootstrap procedure. **B.** Same, for delta band mediated networks, and global efficiency. **C.** Same, for Theta band mediated networks, and local efficiency. **D.** Same, for theta band mediated networks, and global efficiency. **E.** Same, for alpha band mediated networks, and local efficiency. **F.** Same, for alpha band mediated networks, and global efficiency. **G.** Same, for gamma band mediated networks, and local efficiency. **H.** Same, for beta band mediated networks, and global efficiency. Threshold=0.15%, please see Table S1 for p-values.